\documentclass[11pt]{article}
\usepackage{amssymb}
\usepackage{amsmath}
\usepackage[dvips]{graphicx}
\usepackage{color}

\usepackage[T1]{fontenc}

\usepackage{breakcites}  
\usepackage{microtype}   

\usepackage{amsthm}               
\usepackage{amsmath}             
\usepackage{subfig}				

\newtheorem{teo}{Theorem}[section]

\newtheorem{cor}[teo]{Corollary}

\newtheorem{prop}{Proposition}

\newcommand{\be}{\begin{equation}}
\newcommand{\ee}{\end{equation}}
\newcommand{\ba}{\begin{array}}
\newcommand{\ea}{\end{array}}
\newcommand{\bee}{\begin{eqnarray*}}
\newcommand{\eee}{\end{eqnarray*}}
\newcommand{\bea}{\begin{eqnarray}}
\newcommand{\eea}{\end{eqnarray}}

\newcommand{\comment}[1]{}

\newcounter{algo}[section]

\def\IR{\hbox{\rm I\kern-.2em\hbox{\rm R}}}
\def\IC{\hbox{\rm C\kern-.58em{\raise.53ex\hbox{$\scriptscriptstyle|$}}
    \kern-.55em{\raise.53ex\hbox{$\scriptscriptstyle|$}} }}
    
\usepackage{bbm}				
\def \1{{\mathbbm 1}}

\usepackage[autostyle,italian=guillemets]{csquotes}
\usepackage{hyperref}			
\usepackage{authblk}

\title{The financial health of a company and the risk of its default: Back to the future}
  
 \author[1]{Gianmarco Bet\thanks{gianmarco.bet@unifi.it.}}
\author[2]{Francesco Dainelli\thanks{francesco.dainelli@unifi.it. Corresponding author.}}
\author[1]{Eugenio Fabrizi\thanks{eugenio.fabrizi@unifi.it.}}
\affil[1]{\small School of Mathematical, Physical and Natural Science, University of Florence, Viale Morgagni 67, 50134 Florence, Italy. }
\affil[2]{School of Economics and Business, University of Florence, Via delle Pandette 32, 50127 Florence, Italy.}

\date{}

\begin {document}
\maketitle

\begin{abstract}
We theorize the financial health of a company and the risk of its default. A company is financially healthy as long as its equilibrium in the financial system is maintained, which depends on the cost attributable to the probability that equilibrium may decay. The estimate of that probability is based on the credibility and uncertainty of the company's financial forecasts. Accordingly, we develop an equilibrium model establishing ranges of interest rates as a function of predictable corporate performance and of its credit supply conditions. As a result, our model estimates idiosyncratic default risk and provides intrinsically forward-looking PD. \\

\noindent \emph{Declarations of interest:} None.\\

\noindent \emph{JEL classification:} G17, G21, G32, G33.\\

\noindent \emph{Keyword}
financial equilibrium models, probability of default, credit risk, idiosyncratic risk, forward-looking models, credit pricing.

\end{abstract}

\newpage
\section{Introduction}
\label{intro}

When can a company be considered financially healthy? How can we measure and assess the risk (in terms of probability) that this financial health may decline and that the business will default? What contractual power does a borrower have over its lenders? What is the maximum debt a company can sustain in the face of its prospects? How can debt be restructured to restore the company's health? These questions constitute the cornerstones of finance.

	As of yet, models evaluating the default risk of a company and, inversely, its financial health, have failed to produce solid results and their use and their technicalities are disparate in the finance world (corporate finance, credit risk management, financial intermediation, structured finance, project finance, corporate restructuring, etc.). Following several decades of research on the topic, the question of how to measure the probability that a company will fail to meet its contractual obligations on time (Probability of Default, i.e., PD) remains unresolved (\cite{campbell2008search, rajan2015failure, nozawa2017drives}). Clearly, this problem also impacts the issue of how to correctly price the default risk. 
	
	``The failure of models that predict failure" (\cite{rajan2015failure}; see also \cite{hilscher2017credit}) is essentially attributable to two factors. On the one hand, i) {current valuation models underestimate financing constraints (\cite{nikolov2021sources}) and dynamic interdipendent behaviors of operators;}
	on the other hand, ii) soft proprietary information-- usually of a predictive nature-- is neglected in creditworthiness evaluation systems (\cite{agarwal2018loan, gredil2022information}), {where the firm's specific characteristics (asset volatility, growth opportunities, conflict of interests in relation to these estimates, etc.) are fundamental to the scope (\cite{kuehn2014investment})}. For these purposes, the integration of human judgment is necessary in the credit evaluation process\footnote{``Experience and judgment, as well as more objective elements, are critical both in making the credit decision and in assigning internal risk grades." FED - SR 98-25 (SUP) Revised February 26, 2021 - Sound Credit Risk Management and the Use of Internal Credit Risk Ratings at Large Banking Organizations. Since its 2004 revision (known as Basel II), the \cite{basel2006international} also aims for the development of credit evaluation systems integrated with proprietary data elaborated by human judgment. Also, \cite{worldbankgroup2019credit} states: ``The guideline [on credit scoring approaches] encourages the adoption of a human-centric approach, where innovation is applied with the human in mind."}.  
	As a general result, credit ratings have difficulty estimating the borrower's idiosyncratic risk (\cite{hilscher2017credit}). This form of risk is demonstrated to represent the principal element of variations in expected credit loss at the individual level (\cite{nozawa2017drives}), and to be fundamental for pricing objectives (\cite{li2019counterparty}).
	
	Basically, these errors result from models either fitted on data not deriving from any particular theory on the financial health of a company and the risk of its default or are based on an incorrect theory. Models built upon the wrong foundation, or no foundation at all, result in error. The translation of the score produced by these models in a probability of default presents another methodological error, because it is based on frequency analysis of past default behaviors (frequentist approach to probability), instead of looking at the future default drivers of the company in question. In other words, current models study the prediction of a probability instead of the probability of a prediction. 
	
	The use of these faulty models causes severe market inefficiencies (procyclicality, credit crunches, adverse selection, moral hazard, regulatory capital arbitrage). {Moreover, when the regulatory capital of a bank is tightly linked to these models, the prociclicality effects may even be amplified (\cite{becker2014cyclicality}).} This is why there is an increased need for models better able to consider the complexity of the financial market as well as the potentialities of the borrower, reacting to changes in real time, and leading to an evolution in the field of credit scoring\footnote{``Credit scoring has a bright future. There are three potential developments: risk-based pricing [by introducing idiosyncratic risk into evaluations], profitability scoring [by developing systems which look to the future potential of the borrower], and a systems approach that contributes to this bright future [by means of a unitary model of granting, monitoring, renewal, and pricing]," (\cite{johnson2002legal}). }.  
	
	These affirmations are based on the limitations of the main insolvency prediction models that our work aims to overcome, and for this reason, are examined in-depth in the following subsection of this introduction. 
	
	Taking into account the interdependence of operators, another field emerges for potential development, one that is as evident in practice as it is neglected in literature: the estimate of PD made by scoring systems is taken by banks as the basis for recalculating future interest rates to be applied to the borrower. This fact leads to bringing either relief or strain to future debt service (through lower or higher interest rates) and clearly causes a change in the solvency conditions of the counterparty and the relative PD being evaluated. This phenomenon is drastically ignored by current credit evaluation systems. Moreover, that change in PD would in turn produce further rate changes, and so forth.
	
	In conclusion, the true challenge is the development of a theory about the financial health of a company. We need a clear conceptualization of the phenomenon that causes a company's financial distress to begin, develop and erupt in order to analyze and predict the future poor performance of a borrower and to evaluate the possibility (risk) it is not able to meet its obligations on time. Moreover, the challenge lies in elaborating a model that quantifies the probability of that event occurring under a coherent probabilistic framework, where idiosyncratic risk drivers-- evaluated by the use of soft information and human skills-- are fundamental to the scope.
	
	In principle, the financial health of a company depends on maintaining an equilibrium between its demand for and its supply market of credit. It is a function not only of the conditions of the borrower's demand-- the subject of models developed thus far-- but also of the conditions of intermediaries that constitute its credit supply segment (competition, yield curves, availability of information, analysis capacity, etc.). In fact, all of these factors directly influence the ability of the company applying for credit to stay in the market, because they define the availability of resources that are or will be accessible to it and determine at what future cost.
	
	On the basis of this evidence, we form a theory and develop a model which both attempt to meet the challenges outlined above and answer the questions raised in the opening. 
	
	We start from the assumption that a company is financially healthy as long as it is able to maintain equilibrium in the financial system, securing the confidence and interest of those who are asked to grant capital in moments of major need, namely, financial institutions. Equilibrium is lost and a company defaults\footnote{\label{nota}``We distinguish among three concepts associated with the inability to pay" (\cite{bouteille2021handbook}): ``Default is the failure to meet a contractual obligation," meaning that the company is forced to shut down. When a business defaults, it may or may not be in conditions of insolvency, ``which describes the financial state of an obligor whose liabilities exceed its assets." If a business is insolvent, it often goes into ``bankruptcy, which occurs when a court steps in upon default after a company files for protection... of the bankruptcy laws." For purposes of this work, we focus on the concept of default.}  when no lender on the market is willing to provide a loan to that company whose cash inflow inadequately covers its cash outflow (the word ``default" derives from the French ``d\`efaut" meaning lack, shortage or insufficiency). Credit (from the Latin ``creditum": to lend confidence in someone's or something's future potential ability) and the interest of financial operators (from which the ``interest" rate derives, whose magnitude is inversely proportional to the ``interest" shown by lenders) both depend on the credibility (credit-confidence) of the company's forecasts.  The credibility of the business plan in turn depends on the evaluation that the lender makes of the expected solvency{\footnote{{``In dynamic moral hazard models, financing constraints arise from asymmetric information between financiers and insiders. Strictly speaking, such asymmetry of information gives rise to the possibility that insiders ``divert" or blatantly steal cash flows." (\cite{nikolov2021sources})}.}} (from the Latin ``solutum": the ability to ``release", in this case, the binds of debt by means of future payments\footnote{For this reason, expected cash flows are demonstrated to be the most important information in the decomposition of the default risk estimation (\cite{nozawa2017drives, gredil2022information}). }) of the counterparty, which must inspire confidence (not by chance the word ``debtor" derives from the future participle of the Latin ``debere", to owe). Based on the lender's skill at analyzing (sometimes confidential) information about a borrower, the lender makes an assessment of the risk of non-fulfillment of the business plan and of the company's corresponding failure (from the Latin ``fallo": delude, disappoint, deceive, in this case, economically). This assessment is translated into a default probability, which in turn powers the lender's incentive system: together with other risk factors (such as LGD), this system determines the minimum threshold for the fair remuneration of capital lent or to be lent.
	
	While it may seem all sorted out, that is not the case: the establishment of these interest rates changes the numbers of the business plan and thus affects the company's financial health by making future borrowing cheaper or more expensive. This adjustment then alters the PD estimate, which in turn determines new interest rates. This conundrum may be resolved in one of two circumstances: either i) an equilibrium is achieved, reaching a rate at which the PD does not change further. In this case, the company is in equilibrium on the market because it is in good faith with some market operators; or ii) it fails to reach equilibrium because a vicious cycle is set in motion wherein a continually increasing PD estimate would require higher and higher interest rates to compensate for the potential risk. In this case, the company does not earn market trust because the risk does not appear to be adequately rewarded at any rate. 
	
	In order to solve this iterative problem and verify the existence of one or more rates at which the PD does not undergo substantial change, we apply the fixed-point mathematical method. Perhaps surprisingly, we find that this method is only able to reveal a few fixed points, which we consequently consider stable. In order to find the remaining (unstable) fixed points, we resort to the fzero MATLAB routine. The choice of this routine is dictated by numerical constraints.
	
	The main contribution of our study consists of developing a theory about the financial health of a company, based on the maintenance of equilibrium in financial systems characterized by lasting belief manipulation effect in dynamic agency settings{\footnote{{``Our estimates suggest significant financing constraints due to agency frictions and highlight the importance of identifying their sources for firm valuation" (\cite{nikolov2021sources}).}}} with learning and uncertainty, and with rival interdependent principal-agent remuneration systems (\cite{clementi2006theory, cvitanic2013dynamics, he2017optimal}). Under a coherent probabilistic framework-- of a Bayesian interpretation-- a second contribution is the development of a model able to calculate the probability of default and to establish ranges of equilibrium interest rates within which the contractual powers and competitive forces of operators find points of convergence as a function of predictable company performance (variability of cash flow drivers), of changes in its financial structure (leverage intensity, debt maturity structure) and foreseeable trend of its credit supply conditions (rate curves, competition, availability of information, analysis tools, etc.).
	
	The operative result of the model is a credit scoring system that abandons stationary lag settings. The model is based on the use of ``tailored" fundamental analysis techniques\footnote{\label{note5}As early as 1999, the Federal Reserve stated that ``credit risk assessment policies should also properly define the types of analyses to be conducted for particular types of counterparties based on the nature of their risk profile. In addition to customization of fundamental analyses based on industry and business line characteristics, this may entail the need for stress testing and scenario analysis." FED, Supervisory Guidance Regarding Counterparty Credit Risk Management - SR 99-3 (SUP). More recently, see \cite{european2020guidelines}.}.  These are techniques customized on a judgmental basis (\cite{clement2005financial, fracassi2016does}) as a function of multiple competitive and strategic factors/information, and allows analysts (\cite{crane2020skilled, gredil2022information}) and relationship bankers {(\cite{bharath2011lending, brown2021weathering, chava2021impact, han2014informed})} to more accurately evaluate and price risk. Thus, just as in the popular film \emph{Back to the Future}, we imagine returning to the beginning of the 1960s, before the development of ``mass" scoring systems, to divert the history of the financial intermediation market by implementing ``tailored" scoring systems modeled on the rising fundamental analysis techniques.
	
	The development of the theory and model produces further three contributions: i) each company-- or rather, each business plan it presents-- has its own expected default probability, the formulation of the best prediction made at that point, which, therefore, is to be considered fixed for any future moment. Accordingly, the rate regime among different forms of debt depends on exogenous factors to the PD; ii) our model estimates idiosyncratic default risk, accounting for the effect of the expected and unexpected evolution of debt structure over time and introducing human capabilities into the evaluation. As such, it provides an intrinsic forward-looking estimate of the expected portion of PD; and iii) our PD is not independent of demand variables (a higher or lower LGD, for example, allows for rates to be set at higher or lower levels) or of credit supply (especially the interest rate), as these lead to lighter or heavier debt service by means of the ``transmission belt" of interest expenses. 
	
	The remainder of this article is organized as follows: Section~\ref{sec2} theorizes the development of company distress leading to default and conceptualizes the study of its occurrence probability; Section~\ref{model} develops the model; Section~\ref{results} presents and discusses the results, and Section~\ref{conclusion} concludes.

\subsection{Limits of Default Prediction Models} \label{limitsofdefault}

The dominant models in literature and those most widespread among market operators-- primarily the banks (for the U.S. market, see \cite{treacy2000credit})-- are founded on the discriminant-logit-probit analysis (reduced-form models), spread thanks to the regimentation of banking rules as well as the standardization of the credit evaluation processes in larger and larger banks (\cite{stein2002information}). These systems look for commonalities in the distress paths of defaulted companies and trace an ``identikit" by means of a few diagnostic indicators. This early family of reduced-form models is `not founded on a theory of the firm or on any theoretical stochastic processes for leveraged firms" (\cite{crouhy2001prototype}) because it was born under a different paradigm, aimed to provide a score-- not a probability-- to help the credit evaluation process\footnote{\cite{altman1968financial} never speaks about probability, as his original research question relates to ``the quality of ratio analysis as an analytical technique." In fact, he concludes that ``because such important variables as the purpose of the loan, its maturity, the security involved, the deposit status of the applicant, and the particular characteristics of the bank are not explicitly considered in the model, the MDA should probably not be used as the only means of credit evaluation."}.  As a result, reduced-form models that have been rigidly and wrongly employed suffer from important limitations. We explain six of them.
	
	First, they are based on lag measures and not on lead measures. The real drivers of corporate performance and the consequent ability of a company to remain on the market are its competitive and strategic behavior, the true source of idiosyncratic risk. These drivers are summarized by proprietary soft information, which is expensive to acquire, and has an intrinsic prospective nature and value, without which any credit risk assessment system malfunctions~(\cite{rajan2015failure}). Two others follow from this limit: i) these models have a strong predictive capacity in the short term, as they are based on lag indicators, `but this may not be very useful information if it is relevant only in the extremely short run, just as it would not be useful to predict a heart attack by observing a person dropping to the floor clutching his chest" (\cite{campbell2008search}). Therefore, the information is late in the decision-making processes of creditors (this is especially true for indicators based on credit history data, which are obviously the consequence, and not the cause, of the financial stress)\footnote{In the medium term, the predictive power of accounting ratios is demonstrated to be non-negligible and is better than market measures (\cite{campbell2008search}).};  ii) they are intrinsically backward-looking, but the past ``portrait" of a company is often not an accurate indicator of its future plans and strategy, especially in advanced, highly innovative and turbulent economies. In fact, \cite{rajan2015failure} find that ``a statistical model fitted on past data underestimates defaults in a predictable manner." To stem these problems, an attempt was made to take into account the time-series evolution of the predictor variables, using macroeconomic data (\cite{chava2004bankruptcy, das2007common}, and subsequent works). The results at the level of portfolio risk management improved, but the response of the individual borrower to macroeconomic trends (i.e., idiosyncratic risk) continues to be neglected.
	
	Second, they are mainly based on accounting indicators which: i) are erratic, with statistical distributions difficult to manage (improve the predictive power of these models using scaled prices \cite{campbell2008search}), ii) are affected by the accounting environment and accounting policies\footnote{\cite{beaver2005have} find a ``deterioration in the predictive ability of financial ratios for bankruptcy due to increased discretion or the increase in intangible assets not being offset by improvements due to additional FASB."},  and iii) portray a situation which, by the time of the scoring update, has already passed by a few months ``and thus does not fully convey the dynamics of the firm and the continuous process leading to bankruptcy" (\cite{crouhy2001prototype}).
	
	Third, they have margins of error that are not insignificant because: i) each company has its own history and its own decay trajectory which can only partially resemble the statistical ``identikit", and ii) the predictive capacity of many of these models is tested out of sample, even though back-tests are carried out (\cite{coppens2016advances}).
	
	Fourth, they cannot be applied to start-ups.
	
	Fifth, they do not take into account their impact on the company's financial health. In fact, any change in the estimated PD alters the future cost of debt (except for fixed-rate contracts). Consequently, the debt service will be more or less onerous for the borrower. So why shouldn't the PD be affected?
	
	Sixth, they generally work on the total magnitude of the debt position, not taking into account the structure and heterogeneity of debt, which instead influences, often heavily, the debt service charged each year\footnote{\cite{rauh2010capital}: ``We begin by showing the importance of recognizing debt heterogeneity... Studies that treat corporate debt as uniform have ignored this heterogeneity, presumably in the interest of building more tractable theory models or due to a previous lack of data." {On the importance of taking into account the dynamic capital structure, see also (\cite{kuehn2014investment}).}}. 
	
	In addition to these models that power the IRB systems of the vast majority of banks, other models developed by the literature have also come into use. The main reference is to structural models, also used by certain rating agencies (CreditMetrics and KMV). Unlike reduced-form models, structural models are based on a business theory. Here, the default is seen as the event where, on a certain date, the shareholders are no longer interested in continuing the business because the market value of its assets is less than the market value of its liabilities (\cite{merton1974pricing}). Now, the fault is methodological: looking at the capacity of the current value of assets to cover the liabilities, if anything, the (potential) insolvency is evaluated and not the default (see note \ref{nota}). Structural models investigate the wrong phenomenon. As a result, the tools they use are incorrect for running solvency analyses because: i) assets count when they produce cash-flow regardless of their current value, and debt counts when it must be paid\footnote{Future periods with high negative cash flow also may occur when the market value of assets is highly positive-- compatible with the shareholder's ``gamble"-- during which the lenders may decide not to further support the debtor and call back the money. The lender's logic is different from that of the shareholder. },  and ii) there are enormous difficulties in isolating default risk (above all, its idiosyncratic component) by managing market values\footnote{``The equity market has not properly priced distress risk" (\cite{campbell2008search}; see also \cite{nozawa2017drives, chen2022unified}). In our opinion, one cause of this anomaly may lie precisely in the impossibility of stock market measures to isolate, value and even collectivize the great weight of soft information for the purposes of solvency analysis, where idiosyncratic risk plays a fundamental role. In fact, this anomaly was found above all in companies with low analyst coverage and institutional ownership, which are likely difficult to arbitrage.},  as they include many other important factors, not always of a rational nature (\cite{nozawa2017drives, gredil2022information}). 
	
	In the end, the contingent claims approach as well does not provide a proper (default) probability but rather a Distance to Default. From an operative point of view, the biggest limitation of these models is that they can only be applied to a very small group of companies-- those that are listed. Moreover, this approach too suffers from the shortcomings of market-based measures, which are affected by transitory shocks (\cite{gredil2022information}).
	
	Lastly, more recently, machine learning has also made its contribution in this field (\cite{fuster2022predictably} and \cite{sirignano2016deep}). The results are not positive and appear to be generalizable to the field of credit scoring study: a mathematical-statistical technique-- as sophisticated as it may be-- that does not rely on sound corporate theory ``produces predictions with greater variance than a more primitive technology" (\cite{fuster2022predictably}; see also \cite{crouhy2001prototype, rajan2015failure}).
	
	All of these families of models produce scores that need to be translated into a probability. For this purpose, different statistical techniques based on survival analyses are employed (for instance, frequency tables and transition matrices) under a frequentist approach to the study of probability. Considering the structure of our problem (the study of the company's PD largely depends on idiosyncratic drivers\footnote{``We reject the hypothesis that firms' default times are correlated only because their conditional default rates depend on observable and latent systematic factors" (\cite{azizpour2018exploring}).} and exogenous variables slightly controllable at the individual level), the frequentist approach to the study of the single PD is improper. In fact, following this framework, we are finding the probability that the company under evaluation will fail in the future if it were to find itself in similar environmental, competitive and financial conditions experienced by other companies in the past-- even in vastly different sectors-- which subsequently went bankrupt. Clearly, the message we are searching for is different. 
	
	This significant methodological mistake implies the loss of predictive power of individual PD calculated by current models. The term structure of supposed conditional default probabilities is built on this mistake (\cite{jarrow1997markov, duffie2007multi}), as well as on the lifetime PD model which thrives to adopt the recent IFRS 9 (\cite{beygi2018features}). Again, in the end, without a clear conceptualization of the default event based on a correct business financial health theory, not even the occurrence probability of that event can be estimated with a correct predictive value. 
	
	Adding the methodological limitations of the models to the limitations derived by the frequentist approach in the subsequent PD calculation/conversion, stationary and backward-looking rating systems are inexorably generated (see Lucas critique in \cite{mensah1984examination}). As a consequence, current credit origination, monitoring, and pricing processes of financial operators generate inefficiencies in the market, because: i) they are inevitably procyclical (\cite{lowe2002credit}) thus, perhaps making true the famous Mark Twain quote that a ``banker is a fellow who lends you his umbrella when it's sunny and wants it back when it begins to rain." ii) They lead to credit crunches (\cite{behn2016procyclical}), also because startups cannot be evaluated (think of the average lifespan of a high-tech company). iii) ``A blind reliance on statistical default models results in a failure to assess and regulate risks taken by financial institutions" (\cite{rajan2015failure}) and this impoverishes the banking culture because it flattens the skills, competences and information superiorities of the operators and thus lowers the competitiveness of the system. iv) They amplify market inefficiencies because companies that have performed poorly pay more interest (even if they had exceptional business opportunities) and this contributes to worsening their financial situation, starting a vicious cycle that leads to the self-fulfillment of backward-looking systems. v) Measurement criteria of financial assets in the intermediary's balance sheet and stress tests based on these rating systems evidently infects the calculation of the regulatory capital position with the same problems (\cite{behn2016procyclical, cortes2020stress}) {and this tightening may amplify the prociclicality (\cite{becker2014cyclicality})}. In the end, adverse selection and moral hazard phenomena are generated if companies are not and do not feel correctly evaluated. All these factors damage the main transmission belt of monetary policy impulses.

\section{Theory} \label{sec2}
\subsection{Default and the probability of its occurrence} \label{subsec2.1}

Since we are interested in studying default in order to predict it, we need to understand the genesis and the deflagration of this phenomenon. Therefore, the central question is: how and when does a company go into default?

Each company pursues its own strategic plan under the going concern hypothesis, which assumes maintaining support of all its creditors. When the company faces a more difficult scenario compared to its original plan, this causes cash shortages which have an impact on treasury management. If these demands are so burdensome (imagine a serious, extraordinary event) that the company treasury is in severe distress, default can occur immediately. It is very difficult to predict these types of events. More likely, default is a slower trajectory of financial stress-- and therefore, less difficult to predict-- the result of which is a gradual decline in the company's strategic and competitive strength. Focusing on this aspect of default, we estimate the so-called ``expected" portion of PD, or rather, to what extent a company (and its strategic plan) is expected to fail.

In practice, a gradual decline in competitiveness entails greater cash shortages than planned, which, as mentioned above, fall on treasury management. In particular, it is the ``seasonal and the working-capital loans"\footnote{\label{note12}The FED distinguishes ``commercial and industrial loans" in two branches: a) seasonal and working capital loans; and b) term loans. The first types of loans are often structured in the form of an advised line of credit or a revolving credit line. They are short-term facilities that can be flexibly used for a variety of purposes, generally renewed at maturity, periodically reviewed by the bank, without a fixed repayment schedule. See section 2080.12080.1 of Commercial Paper and Other Short-term Uninsured Debt Obligations and Securities - Commercial Bank Examination Manual - FED.}  that provide a business with short-term financing for inventory, receivables, the purchase of supplies, and all other cash needs, including debt service (principal and interest\footnote{\label{note13}``The Company believes its existing balances of cash, cash equivalents and marketable securities, along with commercial paper and other short-term liquidity arrangements, will be sufficient to satisfy its working capital needs, capital asset purchases, dividends, share repurchases, debt repayments and other liquidity requirements associated with its existing operations" - Apple Annual Report 2020, Form 10-K (NASDAQ:AAPL), Published: October 30th, 2020.})  to act as a liquidity buffer and insurance (\cite{kashyap2002banks, gatev2006banks})\footnote{\label{note17}The short-term credit facilities cover the daily unexpected and fluctuating cash needs, both for the nonfundamental component of cash flow volatility (\cite{brown2021weathering}) and for severe financial market disruptions or long-term operational problems (\cite{acharya2014credit, berrospide2015real}).}.  This is aligned with the principles of finance (\cite{shockley1997bank, holmstrom1998private}) that have recently highlighted the benefits of flexibility in short-term facilities and their interest rate structure with respect to term loans\footnote{``We show that without commitment, firms prefer short-term debt for any positive targeted debt financing" \cite{demarzo2021leverage}. }. 

If these cash shortages increase and are not reabsorbed, the corporate distress deepens and credit lines are used more intensively (\cite{ivashina2010bank, campello2012access, brown2021weathering})\footnote{In times of liquidity distress, this way of financing is generally the least demanding in terms of cash needs, since it requires the payment only of interest, and not also of the principal (\cite{campbell2021structuring}).}.  The company reviews its projects baseline according to the severity of the distress, sometimes even radically (for example, changing sales and purchasing policies in certain business areas, foreseeing a recapitalization, changing some managers, etc.). If, once again, the plan faces worse conditions than expected, greater cash shortages are unloaded-- also improperly\footnote{``The following are potential problems associated with working-capital and seasonal loans: \emph{i. working-capital advances used for funding losses}. A business uses advances from a revolving line of credit to fund business losses, including the funding of wages, business expenses, debt service, or any other cost not specifically associated with the intended purpose of the facility." Section 2080.12080.1 of Commercial Bank Examination Manual - FED.}--  on the short-term debt (unless the plan is modified, for instance foreseeing a bond issuance). The rate on these credit lines-- generally variable (\cite{shockley1997bank}) especially for distressed firms (\cite{brown2021weathering})-- increases due to the greater risk perceived by the creditor (contractually, see \cite{shockley1997bank}; for empirical results, see among others \cite{campello2011liquidity}) and this contributes to worsening financial conditions and increasing the cash outflows linked to the debt service itself. Consequently, companies going through liquidity distress cannot help but continue using short-term facilities\footnote{``Distressed borrowers exclusively issue short-term debt," \cite{hu2021theory}.}.  Banks, as a ``liquidity provider of last resort" (\cite{gatev2006banks}), must evaluate whether, and to what extent, to provide additional or backup credit lines. 

If the financial distress continues, credit lines become almost fully drawn down (\cite{luo2020understanding} and \cite{zhao2019usage}) and this sends signals of greater distress to lenders, which, in turn, tend to further increase interest rates and limit access to credit lines (\cite{sufi2009bank, ivashina2010bank, campello2011liquidity, acharya2014credit}). Even medium-long term credit behaves in much the same way (credit crunches and worsening contractual conditions) due to the violation of covenants, further aggravating the company's situation at this critical moment (\cite{sufi2009bank, chodorow2022loan}). 

During such difficult phases when the company and its lenders are deciding whether or not to ``pull the plug", we witness the phenomenon of ``zombie lending" (\cite{hu2021theory}): banks tend to renew short-term credit lines to support cash needs that cannot be postponed-- even when covenants are violated (\cite{campello2011liquidity})-- as long as they believe in a minimum profitability on their investment. Then there comes a time when no bank on the market is willing to grant/extend short-term lines of credit anymore because none believes it to be beneficial at any interest rate, thus, ``when the scheduled payment becomes due and the company does not have enough funds available, it defaults" (\cite{bouteille2021handbook})\footnote{\cite{he2012rollover} demonstrate that the ``credit risk originates from firms' debt rollover" linked to short term maturity, and \cite{sufi2009bank} ``[provides] evidence that lack of access to a line of credit is a more statistically powerful measure of financial constraints than traditional measures used in the literature."}.  The company itself, or a third party, decides that the distress is irreversible and concludes to cease, or demand the cessation of, the activity.

This distress path for any company may unfold in a more or less intense and rapid way, thus every company has its own (higher or lower) probability of default. But there's more. Each strategic plan can lead into default, sooner or later, and with different trajectories, therefore each plan that a company could present to the market has its own probability of entering into default. Therefore, an important conclusion is that the expected PD concerns the probability that the company may fail in carrying out its current plan.

Trying to theorize the distress path described above, we can affirm that the ``expected" default event is the apex of the trajectory of an unplanned chain of events that lead to stress the short-term facilities of a company unable to find an exit strategy out of the distress. Breaking down this definition, default occurs when the company is in the condition:
\begin{enumerate}
\item[a)] of facing unplanned cash flow needs (paying suppliers, employees, interest, etc.);
\item[b)] that no lender is willing to support the company by granting short-term facilities;
\item[c)] that it is unable to develop a credible alternative restructuring plan.
\end{enumerate}

The probability of occurrence of this combined event is very difficult to estimate. Having the company's business plan and future economic and competitive forecasts in hand, points a) and b) above can be forecasted, which is what we will do. What seems impossible to predict are the potential restructuring changes that managers, shareholders, or even third parties (including the State) might make in times of distress-- point c). If this is true, we need to introduce a strong assumption in order to continue: that the company has no possibility of developing alternative plans to the one presented (or rather, any alternative plan cannot be known). Therefore, the ``expected" PD of a borrower is equivalent to the probability of failure of its plan implemented to date, which is the best forecast made as of today of its future financial performance.
Evidently, this assumption leads to a defect in the PD estimate, which often risks being overestimated, even considerably, compared to the actual figure. Luckily, this overestimation is lower for companies already in financial distress, which evidently have fewer paths to take before going default.

\subsection{The forecasting of default} \label{subsec2.2}

The goal is to model the path of decline of a company's financial health under the evolution of its business plan. Specifically, excluding the assumption in point c) referred to in the previous paragraph, it is necessary to verify the manifestation of the other two conditions. The former is a question of simulating the possible onset and deterioration, year by year, of unexpected times of cash flow shortage\footnote{It was found that the main source of agency conflicts in private firms is the cash flow estimate diversion. This diversion drives the investments of a company and the related financial structure (\cite{nikolov2021sources}).} and outlining a variety of distressed trajectories. To do this-- in exercising the principal's monitoring function in a long-term contractual relationship (\cite{clementi2006theory, cvitanic2013dynamics, he2017optimal})--  a financial analyst/operator revises and validates the assumptions on which the plan is based by substantially using fundamental analysis tools\footnote{See note \ref{note5}.}  (\cite{international2007the, chartered2014forward, international2016going, american2017prospective, international2020valuation, international2020business, european2020guidelines}). This review aims to define: 
\begin{itemize}
\item[1)] the levels of bias (accuracy analysis). Prospective financial communication to the market is affected by the classic problems of moral hazard, and there is a risk that it could be characterized by positive bias. The assumptions are revised on the basis of the credibility levels that the analyst attributes to the plan and to its financial projections. 
\item[2)] The degrees of uncertainty (dispersion analysis). The uncertainty of the estimates is quantified in levels of variability that the analyst assigns to the distribution of the probability of verification of all relevant quantities (assumptions), based on publicly available information, confidential information (also depending on the extent and duration of the credit relationship), analytical tools used, skills, and experience.
\end{itemize}

This revision is carried out through the construction and simulation of a variety of hypothetical scenarios on the basis of which analyst consensus and risk estimates are elaborated by means of a plurality of historical analysis tools (correctly based on frequentist approach, which gives the analysis more objective) and soft more or less proprietary information. In line with the definition of default, the simulation generally takes place on the operating and investing cash flows (Free Cash Flow) that the company will produce in the future and on the related financial commitments (debt service)\footnote{\label{note22}The debt service can be integrated with the equity service. In fact, in large companies with a low stock ownership concentration, the payment of dividends can be considered a mandatory cash outflow. It is also true that a company in severe distress generally avoids distributing dividends. The choice is left to the analyst, who can assign levels of bias and degrees of uncertainty to the dividends planned to be paid.}.  The difference between the operating cash flow and the debt service is discharged-- positively or negatively-- on the net short-term financial position.

As for the aforementioned condition b), a credit institution loses interest in financing the company when it estimates that it can no longer exploit a minimal benefit. This occurs when the lender believes that the debtor has entered into irreversible distress and is no longer capable, even in the distant future, of producing sufficient residual cash flow to pay at least fair interest on the debt\footnote{``An obligor is unlikely to pay where interest related to credit obligations is no longer recognized in the income statement of the institution due to the decrease of the credit quality of the obligation." \cite{european2016guidelines}. }.  In fact, in this case the interest would be added to the debt, increasing the losses in the credit relationship over time (growing Exposure At Default). Therefore, default occurs when the bank believes that the short-term facilities are growing irreversibly.

The problem remains of predicting the moment when the distress becomes irreversible. Each business plan consists of an initial analytical forecast period (a period of time necessary for the effects of certain changes explicit in the assumptions to take place) and of a subsequent stabilization of the situation. An event becomes irreversible when it can no longer change state. Therefore, the evaluation of distress irreversibility can only be carried out during the steady state. During the initial period, prolonged and increasing moments of cash shortages do not necessarily mark an irreversible distress but rather can give rise to subsequent moments of solvency. Until the forecast actually reaches a steady state, a determination cannot be made (think of the development of successful giants such as Tesla, which had to go through years of increasing financial needs and debts). Thus, an important conclusion is that distress is irreversible when short-term credit facilities constantly increase in a steady state\footnote{See proof in Proposition~\ref{appendix} in Appendix.}.  This signifies that the company, while executing its business decisions at full capacity, is still unable to repay its financial commitments and thus, the plan fails, demonstrating the failure of investments made and the related financing capital.

In moments of granting, renewing, or monitoring short-term credit lines, the situation is more complicated than described above. In these moments, the default probability is estimated by the lending institution and translated into an affordable interest rate to be requested/applied in the future (in the case of distressed companies, the bank is substantially the price-maker; see \cite{brown2021weathering}). The problem is that this rate alters the estimate of the default probability itself, more or less markedly affecting future cash outflow to service the debt and, therefore, causing the company to decline more or less rapidly towards a situation of irreversible growth of short-term credit facilities. By modifying the PD estimate, the credit pricing also changes again, thus starting a loop. This activates two circles: a vicious circle, when the bank's valuation is detrimental (when the PD estimate raises rates, which in turn raises the PD), and it drains liquidity for the payment of increasing interest; or a virtuous circle, when the valuation is ameliorative, because it allows the borrower to save financial expenses. If the vicious circle does not interrupt itself at a point that keeps the company in equilibrium (i.e., when the worsening of creditworthiness equates to a more than proportional rate increase), it means that there is no rate at which the business plan is sustainable that also satisfies the lender at the same time. Thus, we can better specify that default occurs when an operator believes that there is no interest rate able to cover the estimated cost of the probability of irreversible growth in short-term credit facilities. 

Our model operationalizes this theory. 

This prediction problem, as defined above, is largely founded on idiosyncratic uncertainty drivers defined by the lender, also in function of its credit policy. This kind of data is slightly controllable at the individual level. For these reasons, looking at our phenomenon (largely founded on idiosyncratic uncertainty drivers defined by analysts' skills and by information gathered), we need to consider the partial (lack of) knowledge of the system itself as an essential feature of the model. This ontological vision of the probability corresponds to a Bayesian probabilistic approach, which responds to the following question: what is the probability that the company under evaluation will fail, predicting the trend of the environmental, competitive and financial conditions in which it may find itself in the future?

Developing our theory under this approach, we deduce an important theoretical finding: the probability that the plan will fail must be a single numerical estimate, given that it refers to a single temporally-identified event (short-term debt growth in a steady state). This assumption contrasts with literature and practice that are based on a frequentist approach to probability, which we have already criticized (see subsection~\ref{limitsofdefault} of the previous paragraph). If the probability of the plan default is unique, it must also be the same for every fraction of time during which the plan is carried out before the moment of reaching the steady state. We do not claim there is no one-year PD, rather that probability is the same as that of the best estimate, because any alternative calculation of PD would require additional information unavailable at that time. Evidently, in every successive moment in which the information is updated, the PD can be recalculated. 

If this is true, then there is another important theoretical conclusion: under a certain capital structure planned under specific plan assumptions, the pricing of each form of debt, regardless of the year concession and maturity, is based on the same PD. In other words, if rates depended exclusively on the PD, then all current loans (except those at fixed rates) and all future ones (in any year they were contracted) would have the same rate. We do not assert that pricing is independent of maturity or that a different mix of financial sources have no impact on pricing. In fact, evidently, a different maturity brings contractual characteristics and risk factors (first and foremost, Exposure At Default) that lead the rates to differ. Equally evident is that a different balance between equity/short-term and long-term debt influences the PD-- sometimes even heavily-- because it sizes and distributes cash outflows variously over time, alleviating or worsening the borrower's financial situation.

In this context, the term loans simply constitute a plan assumption, which, like the others, generate their residual effect on the short-term credit lines (see notes~\ref{note12}, \ref{note13} and~\ref{note17}). In compliance with the above theoretical statements, the validation of these assumptions takes place using the same PD, even though the pricing of each term loan can be adjusted for a multiplicity of factors (for example, a different LGD).

Finally, in addition to his own calculations, each operator also thinks about the possible choices of other operators asked to support the plan. In practice, each financial operator tries to predict the solvency analyzes run by other operators interested in the company by reflecting on the information and skills that he/she presumes to possess (for example, a relationship bank rationalizes differently than a new-entry bank). All the work of the financial operator translates into a forecast of interest rates that will be applied on its own credit lines and term loans and on those of other lenders interested in support the borrowing company.

\section{Model} \label{model}

We model long-term credit relationships under asymmetry, uncertainty, signaling, and dynamic learning (\cite{he2017optimal}). Further, we assume rival interdependent incentive systems between the principal and the agent. 

	For simplicity, we consider the financial position financed only by banks, with the same interest rate functions, without constraints or preferences in loan granting, and without any covenants. In other words, as if only one bank were financing the business.
	
	We present the mathematical details of our model. We adopt the notational convention that capital letters indicate random quantities and lower case letters denote fixed quantities. Time is measured in discrete steps $t=0,1,\ldots,T$ (e.g., years). Here $T$ denotes the (random) time when either the debt is paid off or the company defaults. The simulation of treasury management is modeled as follows: ($D_{S,0}=d_{S,0}>0$)
\begin{equation} 
D_{S,t}=D_{S,t-1}-C_t,                                                       
\end{equation}
where $D_{S,t}$ denotes the Short Term Net Financial Position (STNFP) at time $t$, and $C_t=F_t-S_t$ denotes the change in STNFP at time $t$. Here, $F_t$ is the Free Cash Flow generated by operating and investing activities and $S_t$ is the debt served at time $t$\footnote{For simplicity, we do not consider dividend and other equity policies, not even other ancillary cash inflow and outflow (see note \ref{note22}). Nevertheless, it is easy to integrate the proposed model with other variables.}. This is expressed as follows:
\begin{equation}
S_t=c_t+I_{L,t}+I_{S,t}, 
\end{equation}
where $c_t=d_{L,t-1}-d_{L,t}$ denotes the net change in term loans at time $t$ (repayments of term debt net proceeds from issuance of new debt), and $I_{L,t}$ and $I_{S,t}$ respectively represent the interest expense to be paid on the outstanding term debt and STNFP. We assume that interest expense is a linear function of outstanding debt at the beginning of each period: 
\begin{equation}
I_{S,t}=\mathbf r_t D_{S,t}, \qquad       I_{L,t}=\mathbf r_t d_{L,t},                            
\end{equation}
where the $\mathbf r_t$ is the interest rate. To simplify, we assume that $\mathbf r_t=\mathbf r$, for all $t=0,1,\ldots$, in which the rate is constant\footnote{\label{note26}We chose to focus on constant interest rate policies in order to more easily highlight the recursive relationship between rate and default probability. It is possible to extend the class of admissible policies to time-dependent interest rates at the cost of a larger technical overhead and more restrictive assumptions on the underlying dynamics (\cite{adda2002dynamic}). This constitutes an interesting direction for future research.}. Thus, the rate $\mathbf r$ is only a function of the default probability. To emphasize this, we write $\mathbf r=\mathbf r(p)$.

	In order to model the impacts of analyst's revisions, we assume that $F_t$ is a random variable with mean (bias) $\mu$ and standard deviation $\sigma^2$. For simplicity's sake, we take $\mu$ and $\sigma^2$ to be independent of $t=0,1,\ldots$ but our analysis can be easily extended to time-dependent forecasts. Thus, $\mu$ and $\sigma^2$ respectively represent the reliability of the plan and the uncertainty of the plan according to the analyst's revisions. Crucially, in accordance with a Bayesian probabilistic framework, they are input parameters that can be fine-tuned by the analyst. We denote by $F_t (\omega)$ an outcome of the random variable $F_t$, where $\omega$ denotes the sample. We assume that a company enters into a steady state after a certain time $t_{SS}$, where $F_t=F$ is constant for $t\geq t_{SS}$.

	In order to estimate the mean and distribution of various quantities of interest, we simulate the underlying random process a large number of times and then take empirical averages. Let us illustrate this point and explain how we estimate the default probability. We define the default event as an increase of STNFP in a steady state (see Proposition~\ref{appendix} in Appendix). Formally, the default event is 
\begin{equation}
\mathcal D^i\colon = \{\exists t>t_{SS}  \text{ such that }  C_t (\omega_i )=F_t (\omega_i )-S_t (\omega_i )<0\},                         
\end{equation}
where $i=1,\ldots,N$ denotes the specific sample. The default probability $\mathrm{PD}$ is approximated as
\begin{equation} \label{pdfun}
\mathrm{PD}\approx \overline {\mathrm {PD}}\colon = \frac{1}{N} \sum_{i=1}^N \mathbf{1}(\mathcal D^i ) =\frac{\sharp\{{\omega_i:C_t (\omega_i)<0,t>t_{SS}}\}}{N}.                                  
\end{equation}
When $N$ is large enough, $\mathrm{PD}$ does not depend on the samples $\omega_1,\ldots,\omega_N$ and thus we omit it from the notation. On the other hand, the default probability depends crucially on the interest rate $r$ and we emphasize this in the notation as $PD=PD(r)$. In general, if $r\in(0,1)$ is some rate, and we set $p=PD(r)$, then $\mathbf r(p)\neq r$. This motivates our definition of the equilibrium rate. To this end, we define the composite function
\begin{equation}
\tau(r)\colon =\mathbf r(PD(r)).                                                       
\end{equation}

	The equilibrium rate $r_{eq}$ is the fixed point of the function $r\rightarrow \tau(r)$, that is, $r_{eq}$ satisfies 
\begin{equation}
\tau(r_{eq} )=r_{eq}.         
\end{equation}                                
Computing $r_{eq}$ explicitly is, in general, not straightforward, and thus we turn to approximation techniques. Considering the recursive structure of the problem, we choose a robust technique known as the fixed-point method (\cite{burden2015numerical}). In short, this method generates a sequence $r_k$ such that $r_k=\tau (r_{k-1} )$, and then approximates $r_{eq}\approx r_k$ for $k\gg1$. However, the fixed-point algorithm only retrieves a subset of fixed points, which we therefore consider stable. In order to compute the unstable fixed points, we resort to the more sophisticated \emph{fzero} MATLAB routine. This uses a combination bisection, secant, and inverse quadratic interpolation methods (\cite{brent2013algorithms, forsythe1977computer}).

	In order to better study the behavior of the lender, we relate the equilibrium rate defined above to the rate that maximizes the benefit for the bank. The bank profitability is defined as
\begin{equation}
\begin{split}
R(\omega)=&\sum_{t=1}^T  \frac{C_t (\omega)+I_{S,t} (\omega)+c_t+I_{L,t}}{\alpha^t} - d_{S,0}- d_{L,0}+\\
&+\left[\frac{D_{S,T} (1-LGD)}{\alpha^t} +\frac{d_{L,T} (1-LGD)}{\alpha^t} \right]\mathbf{1}(\mathcal D^\omega ),             
\end{split}
\end{equation}
if $D_{S,T}\leq d_{S,0}$, or
\begin{equation}
\begin{split}
R(\omega)=&\sum_{t=1}^T  \frac{C_t (\omega)+I_{S,t} (\omega)+c_t+I_{L,t}}{\alpha^t} - d_{S,0}- d_{L,0}+\\
&+\left[-\frac{D_{S,T}-d_{S,0}}{\alpha^t} +\frac{d_{S,0} (1-LGD)}{\alpha^t} +\frac{d_{L,T} (1-LGD)}{\alpha^t} \right]\mathbf1(\mathcal D^\omega ),
\end{split}
\end{equation}
if $D_{S,T}>d_{S,0}$\footnote{We assume LDG equals 1 for the amount of debt exceeding $d_{S,0}$, since we can assume that credit guarantees are estimated by the bank to cover only the present debt. }.  \\
The constant $\alpha >1$ is a discount factor and LGD is the so-called Loss Given at Default. 
We estimate the expected return as
\begin{equation}
\bar R\approx \frac{1}{N} \sum_{i=1}^N R(\omega_i ) ,                                                          
\end{equation}
where, again, we are allowed to drop the dependence on $\omega_1,\ldots, \omega_N$ if $N$ is large enough. On the other hand, we emphasize the dependence of $\bar R$ on the interest rate by writing $\bar R=\bar R (r)$.

\section{Results} \label{results}

We numerically simulate a growing company, assuming it reaches a steady state after 5 years, formally $t_{SS} =5$. Following an indirect method, the Free Cash Flow $F_t$ is broken down into its main components in order to better study its variability as a function of the analyst's revision, formally
\begin{equation} \label{flusso}
FCF = F_t = Rev_t - C_t^{var} - C_t^{fix} - Tax_t + C_t^{WC} - Cap_t,                         
\end{equation}
	Table~\ref{table1} presents data for our first simulation (\textit{Case A}).

\begin{table}[ht]
\centering
\caption{Input parameters for FCF simulation (Case A). This table reports the input parameters for our FCF simulation as in equation (\ref{flusso}) for Case A. $x$ indicates the percentage of the corresponding variable planned by the company while $\mu$ and $\sigma^2$ respectively denote the mean and variance of the corresponding random noise.}
\label{table1}
\begin{tabular}{|l|c|c|c|c|}
\hline
 & $x$ & $\mu$ & $\sigma^2$ & Formula \\
\hline
Revenue $Rev_t$ & $x_{Rev}=10\%$ & $-0.10$ & $0.10$ & $Rev_{t-1} (1+x_{Rev}+\varepsilon_{Rev,t} ) $  \\
&&&& with $Rev_0=3000$\\
\hline
Variable Cost $C_t^{var}$ &  $x_{var}=30\%$ & $0.05$ & $0.02$ & $Rev_t (x_{var}-\varepsilon_{var,t} )$ \\
\hline
Fixed Cost $C_t^{fix}$ & $-$ & $0.05$ & $0.01$ & $400(1+\varepsilon_{fix,t} )$ \\
\hline
Tax $Tax_t$ & $x_{Tax}=30\%$ & $-$ & $-$ & $\max \{0,(Rev_t-C_t^{var}-C_t^{fix} ) x_{Tax} \}$ \\
\hline
Change in NWC $C_t^{WC}$ & $x_{WC}=1\%$ & $-$ & $-$ & $x_{WC} Rev_t$ \\
\hline
Capex $Cap_t$ & $-$ & $0.05$ & $0.01$ & $40(1+\varepsilon_{Cap,t} )$ \\
\hline
\end{tabular}
\end{table}

\begin{figure}[ht]
  \centering
  \includegraphics[scale=0.5, trim=30mm 80mm 30mm 80mm\textwidth]{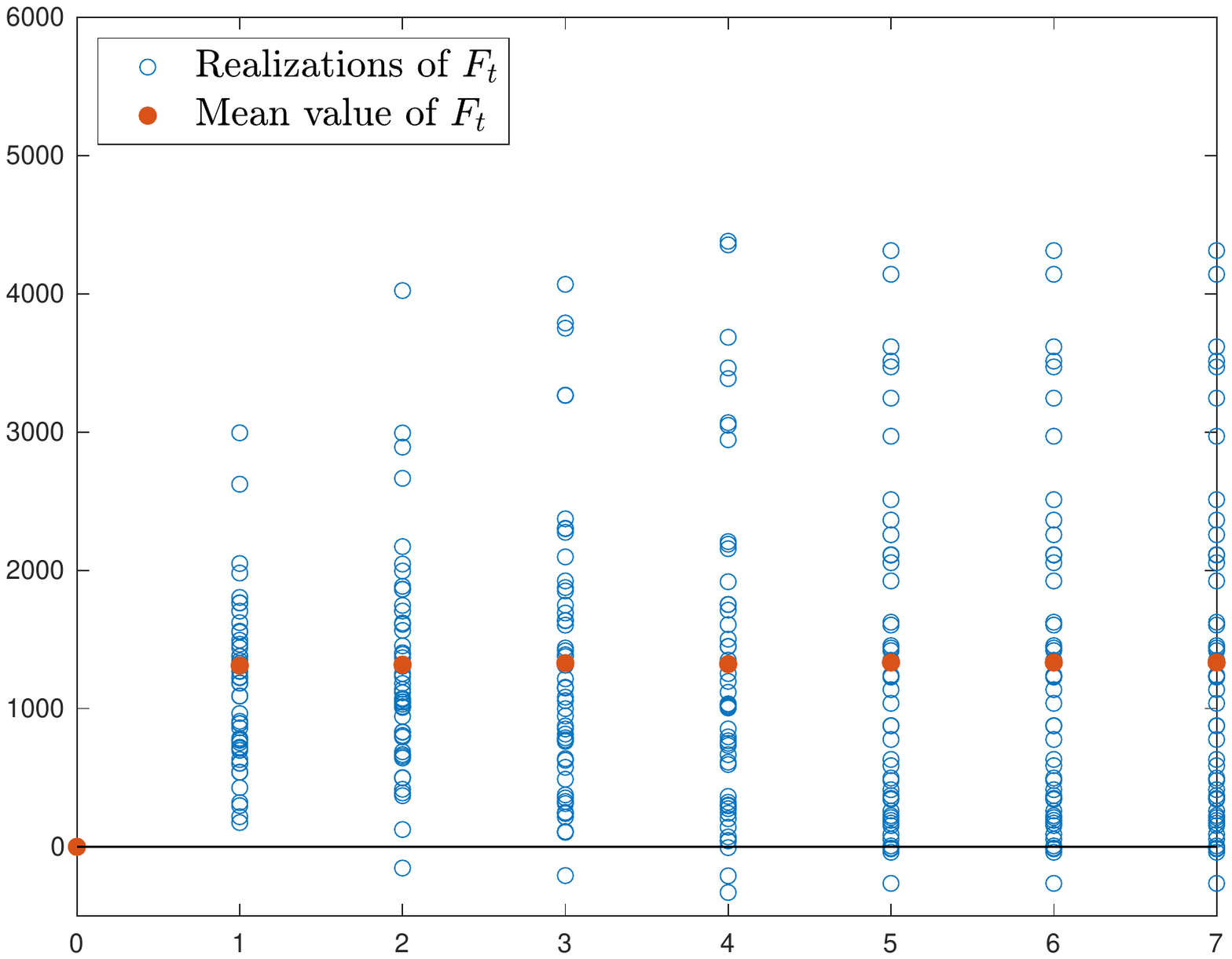}
  \caption{The FCF simulation results for Case A. 50 realizations of $F_t$, in blue. The mean value (in red) is calculated on 2500 realizations of $F_t$.}
  \label{fig1}
  \end{figure}

The results of the simulation are shown in Figure~\ref{fig1}.

  In this scenario, the bias introduced by the analyst (i.e., the mean of $\varepsilon_{Rev,t}$) cancels out the growth anticipated in the plan (i.e., $x_{Rev}$). Accordingly, the most plausible scenario approximated by the distribution average stabilizes in the five-year period around the FCF declared by the company. 
  
	In order to respond to the first three questions posed in the introduction, we develop two more cases.
	
	\textit{Case B}. In this scenario, the variance of the error terms is halved, as shown in Table~\ref{table2}.

  \begin{table}[ht]
\centering
\caption{Mean and variance of random noise for FCF simulation (Case B).
This table reports the mean and variance of random noise for our FCF simulation as in equation (\ref{flusso}) for Case B. The variance is halved compared to Case A, while other input parameters remain unchanged.}
\label{table2}
\begin{tabular}{|l|c|c|}
\hline
 &  $\mu$ & $\sigma^2$   \\
\hline
Revenue $Rev_t$ & $-0.10$ & $0.05$ \\
\hline
Variable Cost $C_t^{var}$ & $0.05$ & $0.01$  \\
\hline
Fixed Cost $C_t^{fix}$ & $0.05$ & $0.005$  \\
\hline
Capex $Cap_t$ & $0.05$ & $0.005$  \\
\hline
\end{tabular}
\end{table}

\begin{figure}[ht]
  \centering
  \includegraphics[scale=0.5, trim=30mm 80mm 30mm 80mm\textwidth]{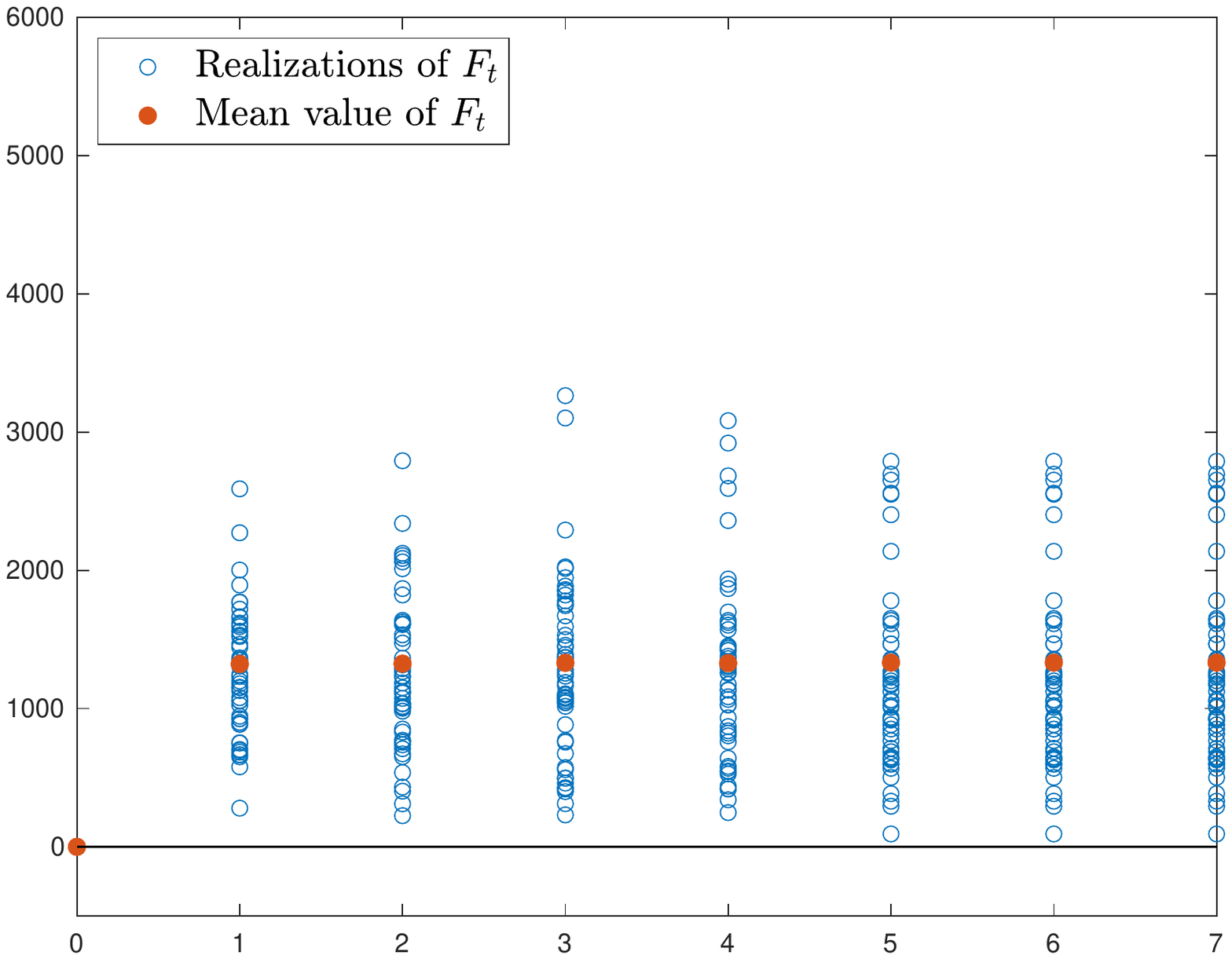}
  \caption{The FCF simulation results for Case B. 50 realizations of $F_t$, in blue. The mean value (in red) is calculated on 2500 realizations of $F_t$.}
  \label{fig2}
  \end{figure}

Accordingly, the resulting flow is sharply concentrated around its mean (Figure~\ref{fig2}).

\textit{Case C}. In this scenario, we reduce the bias as shown in Table~\ref{table3}.   

  \begin{table}[ht]
\centering
\caption{Mean and variance of random noise for FCF simulation (Case C).
This table reports the mean and variance of random noise for our FCF simulation as in equation (\ref{flusso}) for Case C. The mean is halved compared to Case A, while other input parameters remain unchanged.}
\label{table3}
\begin{tabular}{|l|c|c|}
\hline
 &  $\mu$ & $\sigma^2$   \\
\hline
Revenue $Rev_t$ & $-0.05$ & $0.10$ \\
\hline
Variable Cost $C_t^{var}$ & $0.025$ & $0.02$  \\
\hline
Fixed Cost $C_t^{fix}$ & $0.025$ & $0.01$  \\
\hline
Capex $Cap_t$ & $0.025$ & $0.01$  \\
\hline
\end{tabular}
\end{table}

\begin{figure}[ht]
  \centering
  \includegraphics[scale=0.5, trim=30mm 80mm 30mm 80mm\textwidth]{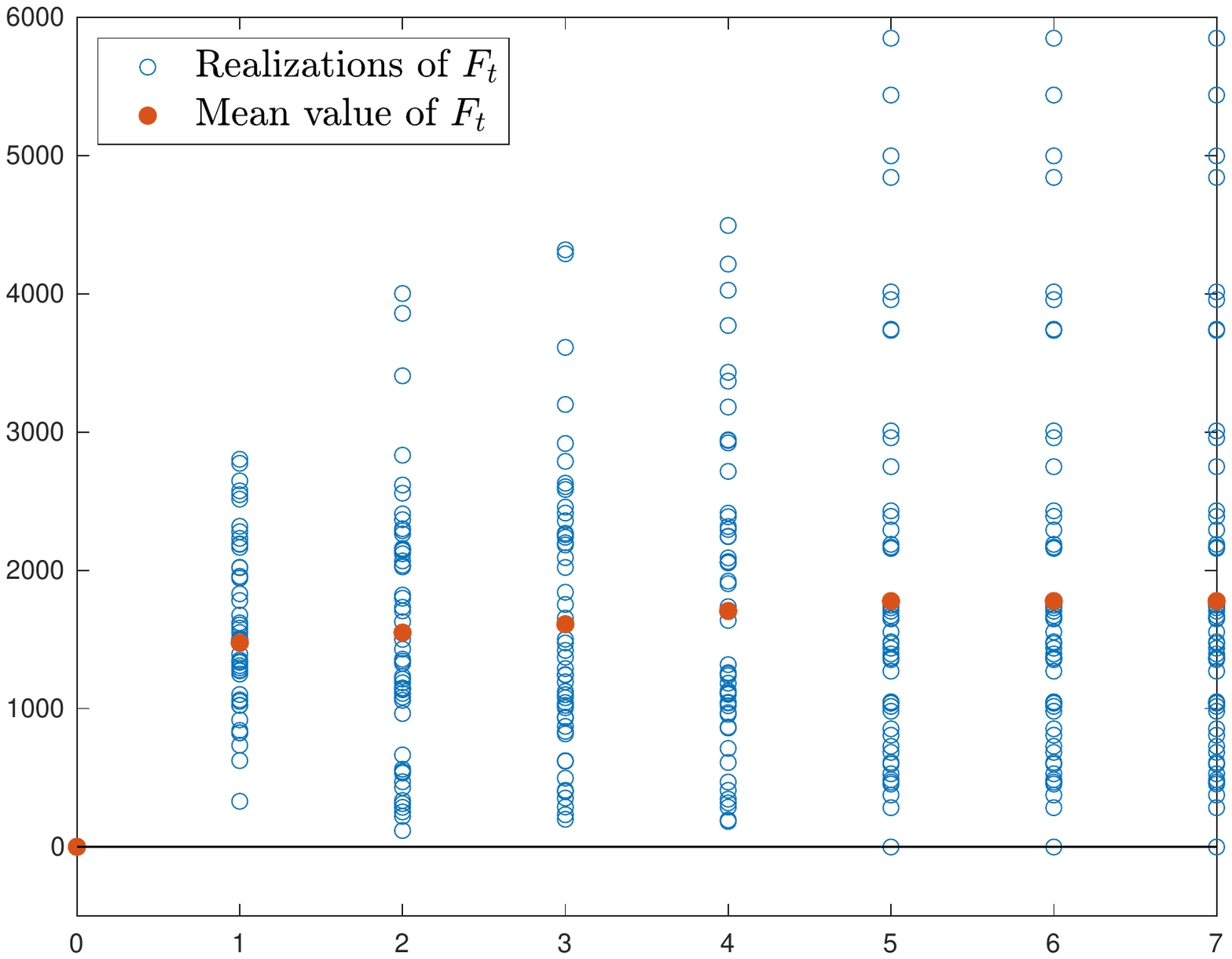}
  \caption{The FCF simulation results for Case B. 50 realizations of $F_t$, in blue. The mean value (in red) is calculated on 2500 realizations of $F_t$.}
  \label{fig3}
  \end{figure}

Accordingly, the resulting flow increases on average, as shown in Figure~\ref{fig3}.

\begin{figure}[]
  \centering
\makebox[\textwidth][c]{
\subfloat[] 
	{	\includegraphics[scale=0.5, trim=30mm 80mm 30mm 80mm\textwidth]{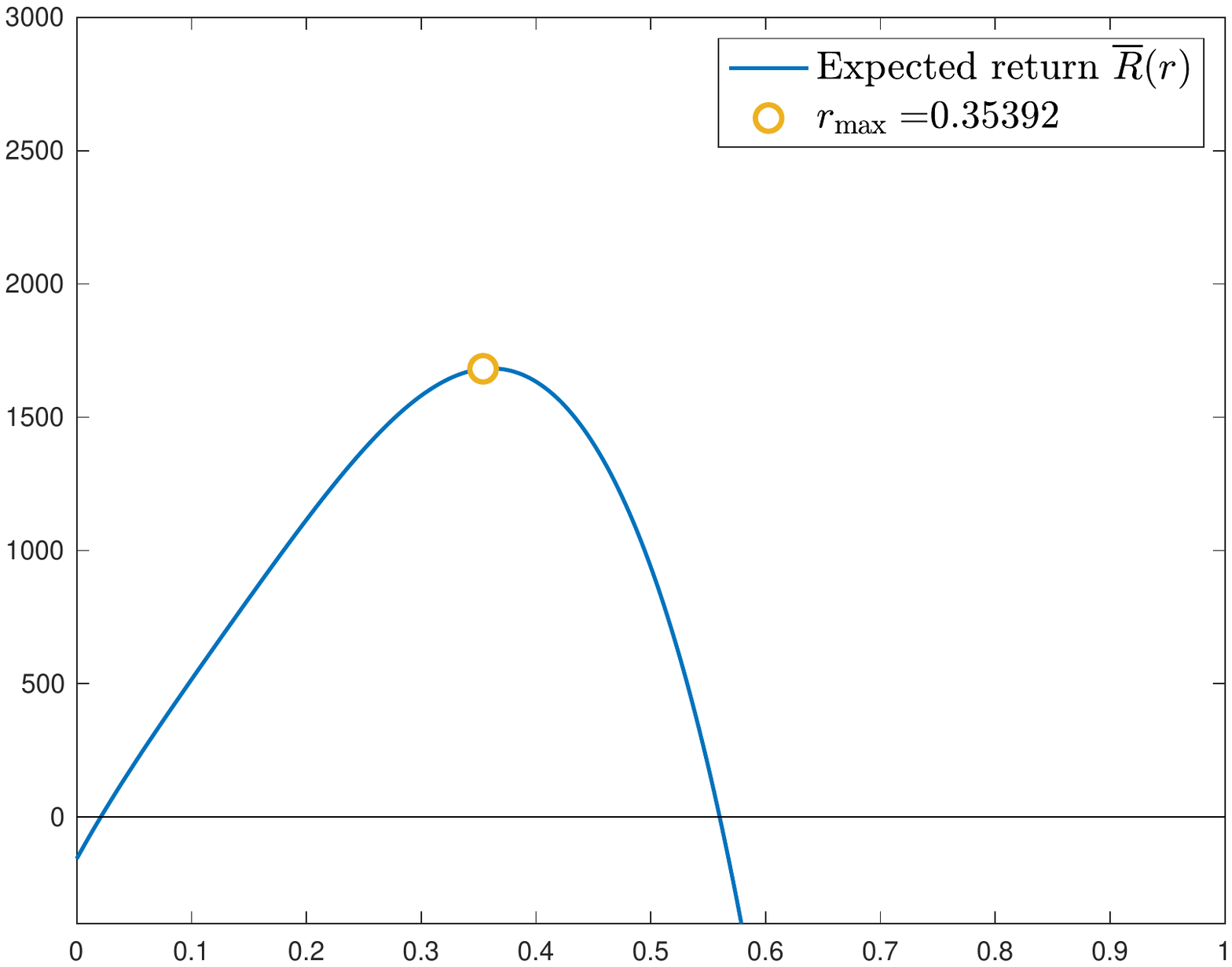} \label{fig4a}} 
	\hspace{1.4cm}
\subfloat[]
	{	\includegraphics[scale=0.5, trim=30mm 80mm 30mm 80mm\textwidth]{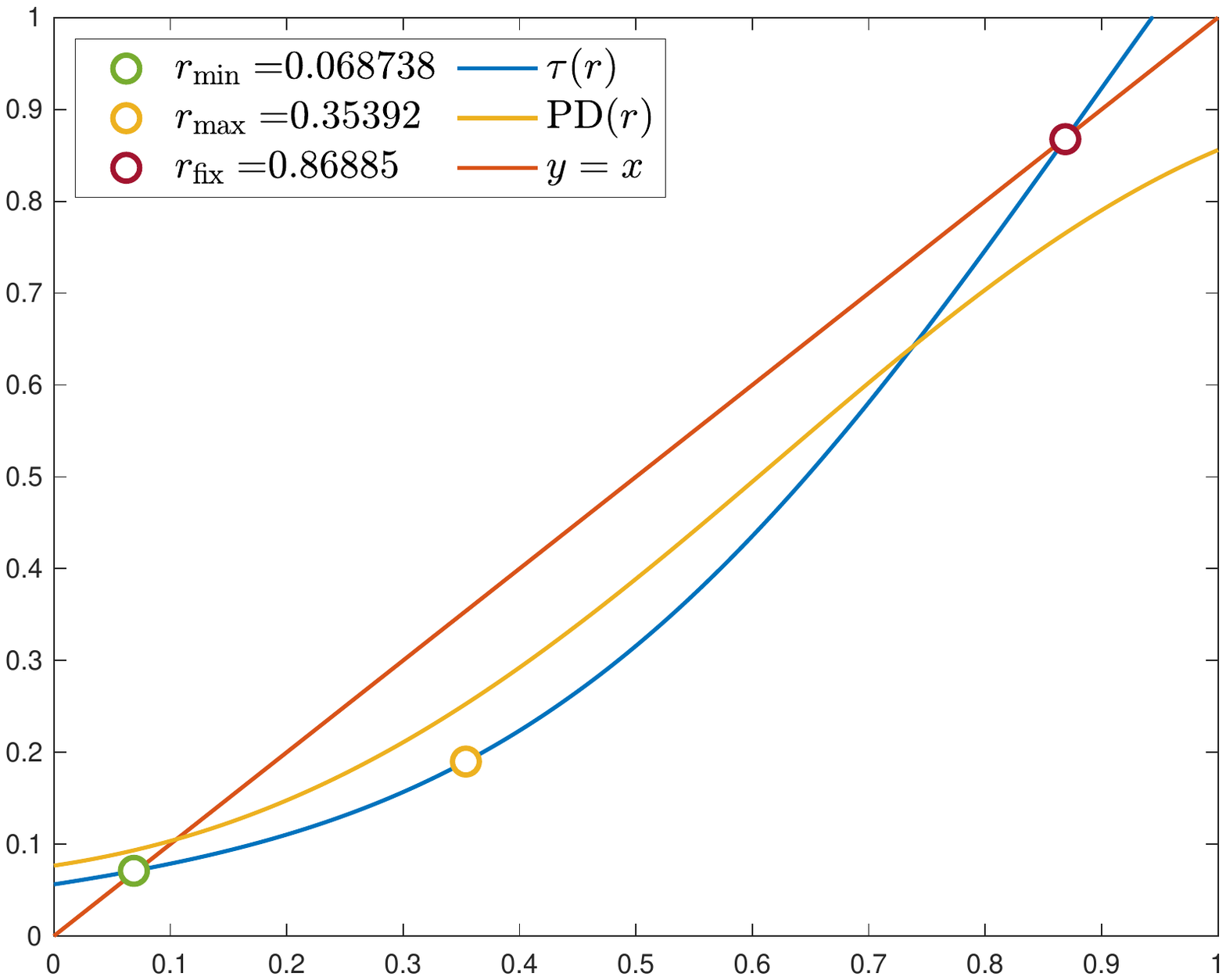} \label{fig4b}}
	} \\
\subfloat[]
	{	\includegraphics[scale=0.7, trim=0mm 0mm 0mm 0mm\textwidth]{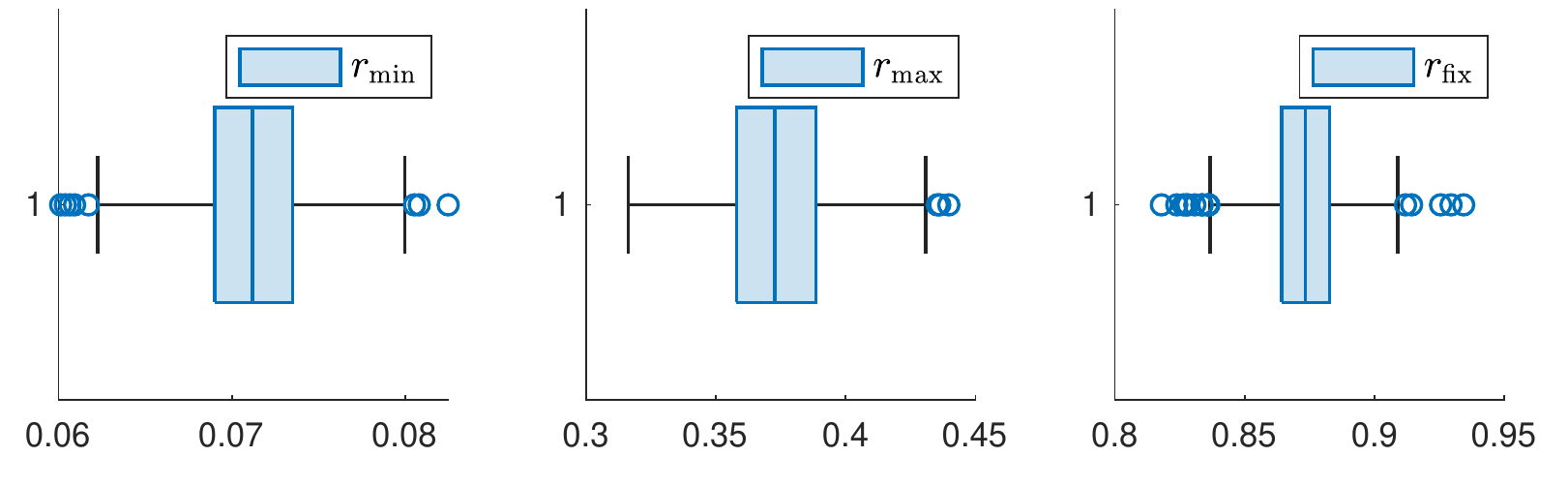} \label{fig4c}}
\caption[]{Simulation results for Case A with $N=2500$ samples. Figure~\ref{fig4b}: function $PD(r)$ in yellow, resulting from the PD function (\ref{pdfun}) for each value of $r\in(0,1)$. Function $\tau(r)$ in blue represents the rate value $\tau$ that the bank should apply based on the PD induced by rate $r$. Line $y=x$ is in red. The intersection between line $y=x$ and function $\tau$ are fixed points of $\tau$: $r_{\min}$ in green and $r_{\mathrm{fix}}$ in red. The results are restricted to the space $(0,1)$, beyond which the method detects other fixed points that are unrealistic. Figure~\ref{fig4a}: Expected return $\bar R (r)$ for the lender for each applicable interest rate. The yellow point $r_{\max}$ is the maximum expected return with discount rate $\alpha=1.01$. This same point has been reported in the graph at the top right. Figure~\ref{fig4c}: Boxplot of our estimators for $r_{\min}$, $r_{\max}$, and $r_{\mathrm{fix}}$ based on 1000 point estimates. }
\label{fig4}
\end{figure}

As for the debt service, we consider:
\begin{enumerate}
\item STNFP $d_{S,0}=2000$.
\item Term debt 1000, issued in $t=1$, to pay back within 10 years starting from $t=2$ with an even principal payment schedule.
\item Rate: $\mathbf r(p)=\frac{r_f+p\times LGD}{1-p\times LGD}$ where $r_f=0.01$ constant, $LGD=0.6$ constant. 
\end{enumerate}
The simulation results for Case A are shown in Figure~\ref{fig4}. 

The boxplots of our estimators for $r_{\min}$, $r_{\max}$ and $r_{\mathrm{fix}}$ highlight that, even with our conservative choice of $N=2500$ samples, the estimators are fairly accurate.

	In this case we find two fixed points. The first ($r_{\min}$) corresponds to a stable fixed point identified by the fixed-point method. Technically, the fixed-point method is successful because the absolute value of the derivative of $\tau$ in $r_{\min}$ is less than 1, that is, $|\tau' (r_{\min} )|<1$ (\cite{burden2015numerical}). This means that for any rate less than $r_{\min}$, the model converges toward $r_{\min}$. Similarly, this convergence is observed for each point greater than $r_{\min}$, up to the fixed point called $r_{\mathrm{fix}}$. The point $r_{\mathrm{fix}}$ is an unstable fixed point (the absolute value of the derivative of $\tau$ in $r_{\mathrm{fix}}$ is greater than 1, i.e., $|\tau' (r_{\mathrm{fix}} )|>1$). We obtained $r_{\mathrm{fix}}$ by means of the \emph{fzero} MATLAB routine, since applying the fixed-point method to a starting point larger than $r_{\mathrm{fix}}$ results in diverging rates.
	
	The significance of these results is important. Any rate less than $r_{\min}$  is not remunerative for the lender's risk and would need to be increased (since the derivative of $\tau$ is less than one, this means that the application of a rate $x$ generates a certain PD, which would result in a rate $y$ greater than $x$). Each rate between point $r_{\min}$ and $r_{\mathrm{fix}}$ corresponds to an over-remuneration of the lender, a situation that offers the borrower room for negotiation to reduce rates. For points beyond $r_{\mathrm{fix}}$, the financial distress on the borrower's finances provoked by such high rates cannot adequately cover the lender for the interest rate, which would require it to be progressively increased. However, it's in the lender's best interest to stop well before $r_{\mathrm{fix}}$. In fact, he or she has no interest in going beyond point $r_{\max}$. Beyond this point, the expected return tends to decrease as the default risk caused by applied rate hikes increases (see Figure~\ref{fig4a}). In conclusion, in partially efficient markets, rate bargaining between borrower and lender takes place between point $r_{\min}$ and point $r_{\max}$. 
	
\begin{figure}[]
  \centering
\makebox[\textwidth][c]{
\subfloat[] 
	{	\includegraphics[scale=0.5, trim=30mm 80mm 30mm 80mm\textwidth]{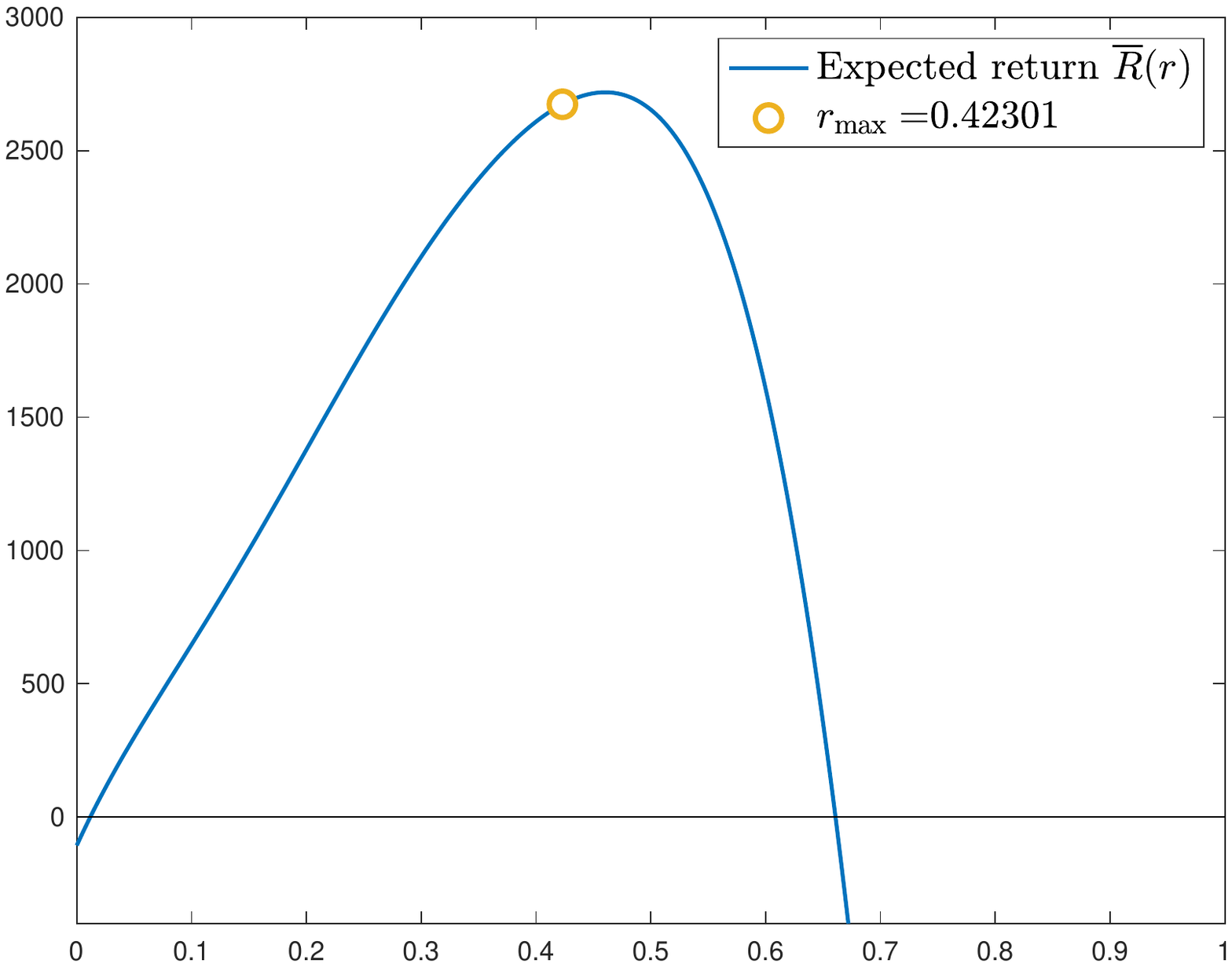} \label{fig5a}} 
	\hspace{1.4cm}
\subfloat[]
	{	\includegraphics[scale=0.5, trim=30mm 80mm 30mm 80mm\textwidth]{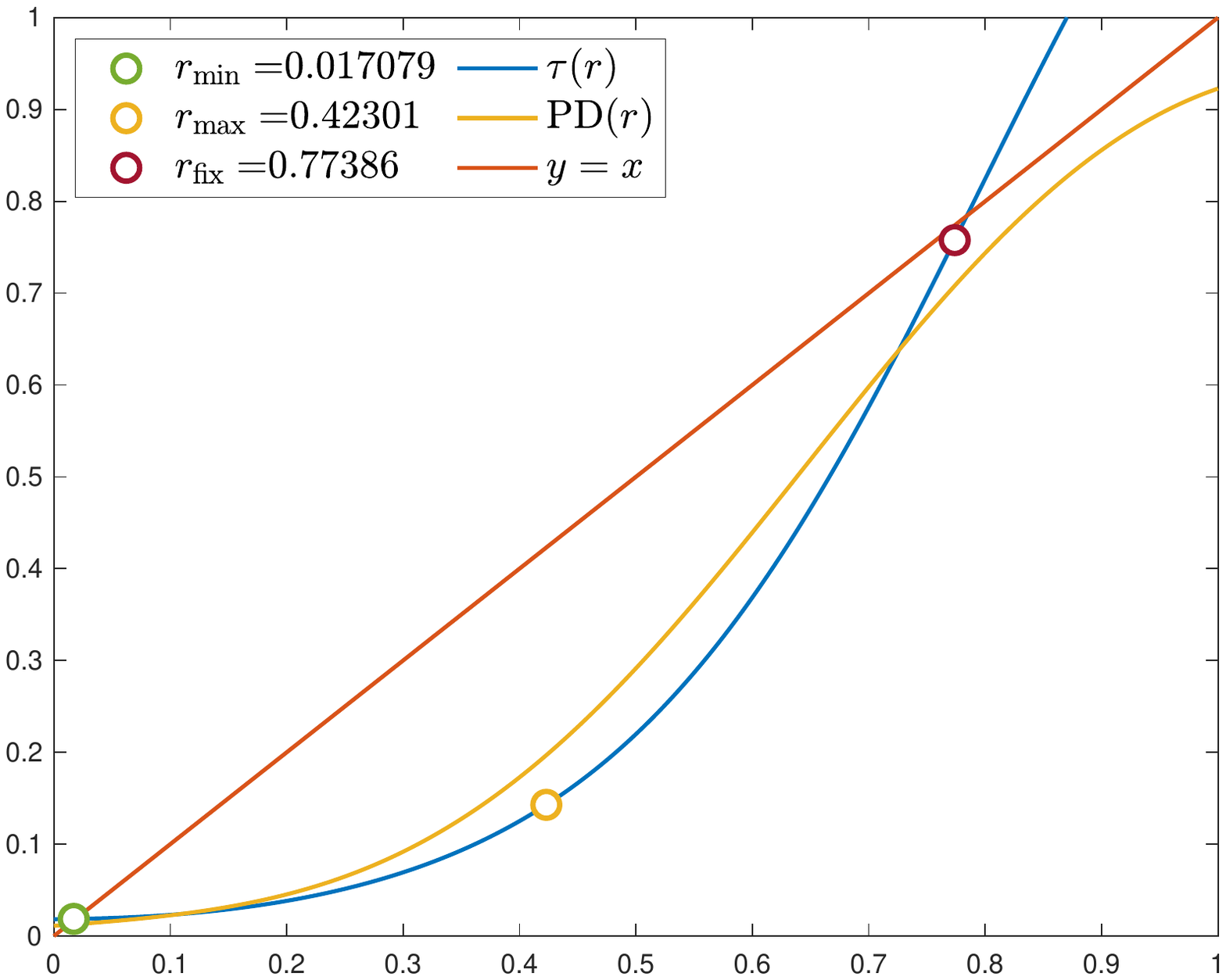} \label{fig5b}}
	} 
\\
\subfloat[]
	{	\includegraphics[scale=0.7, trim=0mm 0mm 0mm 0mm\textwidth]{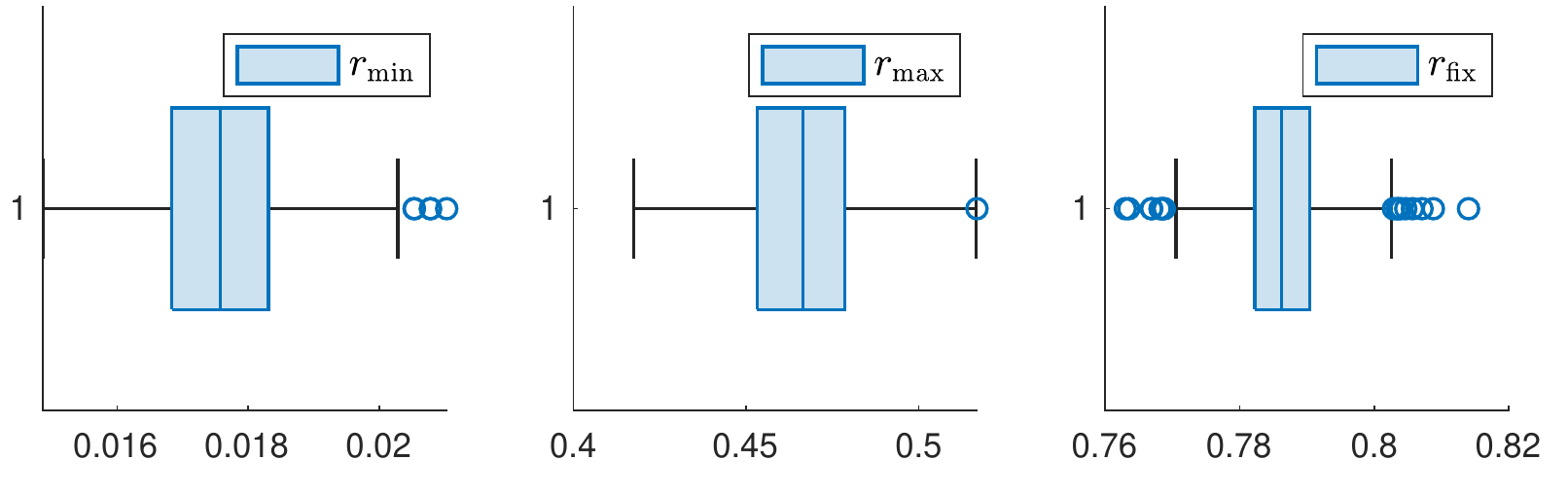} \label{fig5c}}
\caption[]{Simulation results for Case B. }
\label{fig5}
\end{figure}

	In Case B (see Table~\ref{table2}), the reduction in variance evidently reduces the risk that is assessed by the creditor (Figure~\ref{fig5}).

It follows that point $r_{\min}$ moves correctly on rates close to $r_f$. Even higher rates (up to $r_{\max}$) overwhelmingly converge to $r_f$. This happens because the rate of convergence of the methods depends on the value of the absolute value of the derivative of $\tau$ in $r_{\min}$. The smaller the value of $|\tau' (r_{\min} )|$, the faster the convergence, which may be very slow if $|\tau' (r_{\min} )|$ is close to 1 (\cite{burden2015numerical}). This means that, by reducing the uncertainty of the forecast, the risk of default is so low that the bank's negotiating power should be reduced in favor of that of the borrower compared to the base scenario. Evidently, by reducing the uncertainty in the data, when the rate (after the $r_{\max}$) begins to be unsustainable against business cash flow, it moves towards a situation of failure much faster than in the base scenario. In conclusion, the curvature of the blue line close to $r_{\min}$ (i.e., the second derivative $\tau'' (r_{\min})$) determines the negotiating power of the borrower and the lender. The flatter the curve around $r_{\min}$ (i.e., the closer $|\tau' (r_{\min} )|$ to zero), the more power is in the hands of the company. 

\begin{figure}[]
  \centering
\makebox[\textwidth][c]{
\subfloat[] 
	{	\includegraphics[scale=0.5, trim=30mm 80mm 30mm 80mm\textwidth]{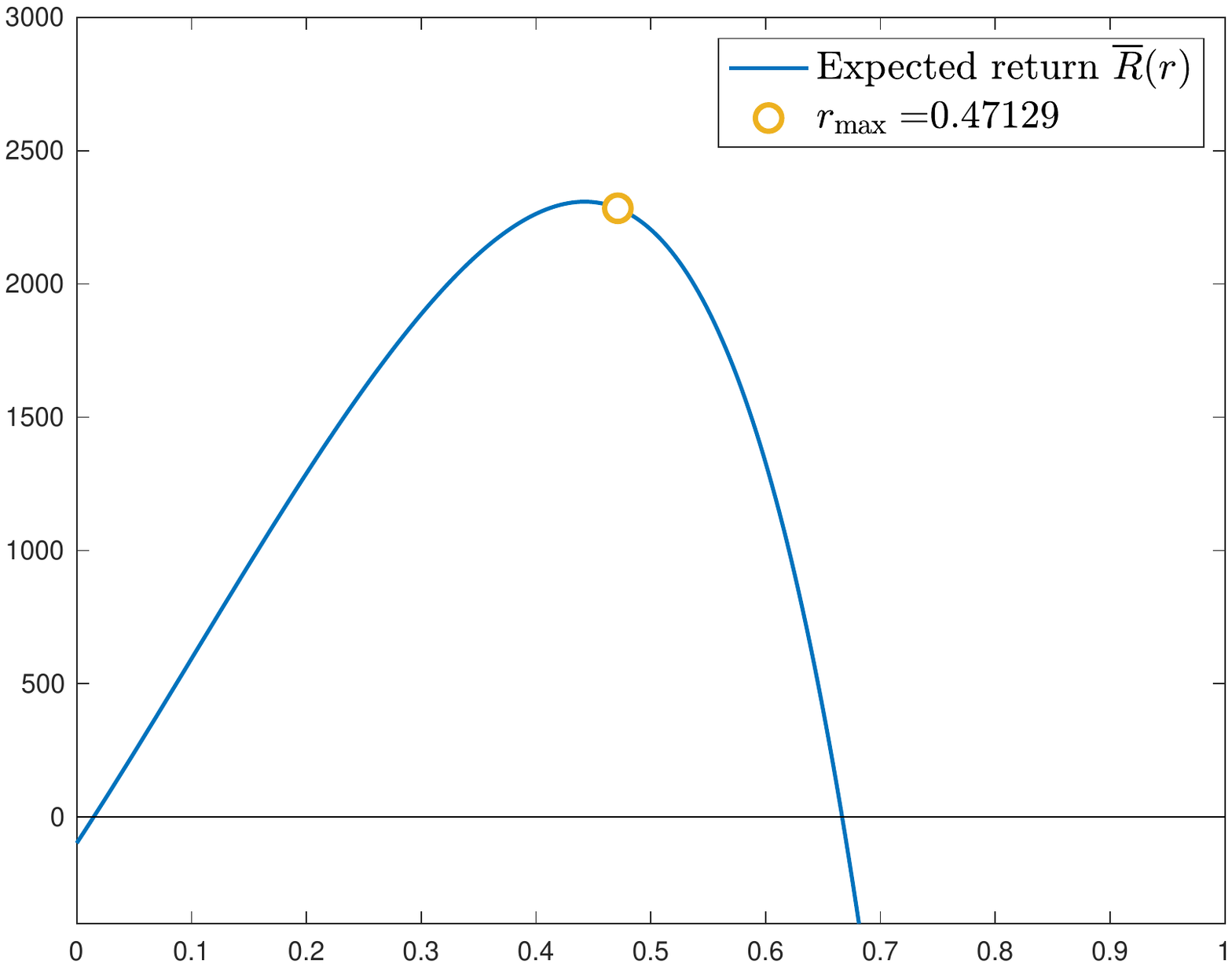} \label{fig6a}} 
	\hspace{1.4cm}
\subfloat[]
	{	\includegraphics[scale=0.5, trim=30mm 80mm 30mm 80mm\textwidth]{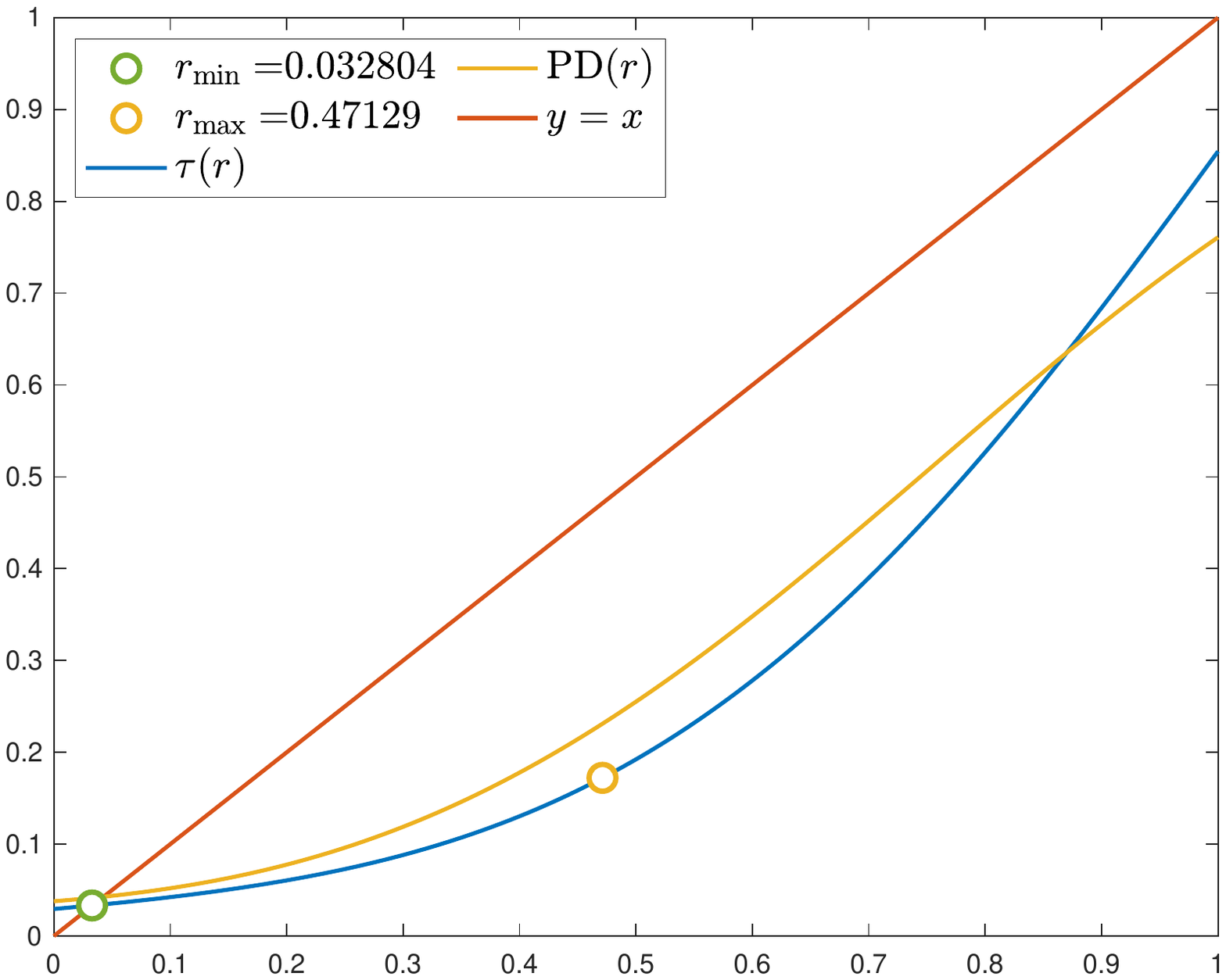} \label{fig6b}}
	} 
\\
\subfloat[]
	{	\includegraphics[scale=0.7, trim=0mm 0mm 0mm 0mm\textwidth]{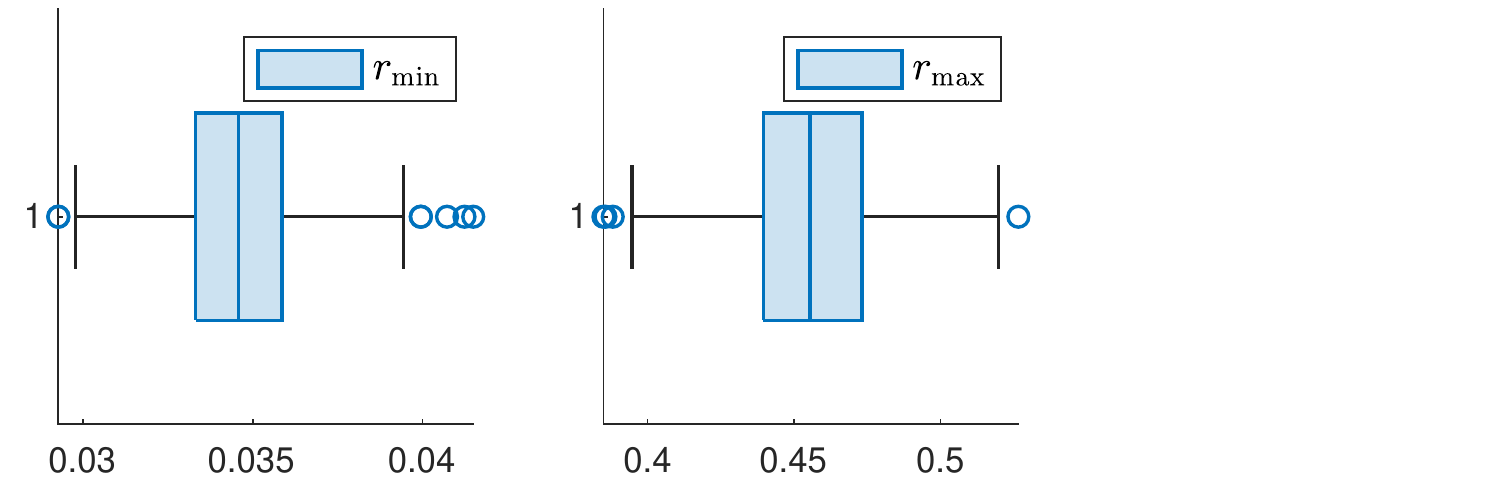} \label{fig6c}}
\caption[]{Simulation results for Case C.  }
\label{fig6}
\end{figure}

	In Case C (see Table~\ref{table3}), the reduction of the bias gives greater credibility to the verification of the company's planned scenario, which claims to have growing cash flow (Figure~\ref{fig3}). It simply follows a shift of the blue curve $\tau$ to the right (Figure~\ref{fig6b}). The credit institution realizes that it has a customer that is easier to finance in its hands, even at lower rates, compared to what is estimated in the base scenario (Figure~\ref{fig4}).

Our findings are confirmed in the literature and offer an interpretative line to its results. First of all, our model shows to what extent the size and cost of credit lines are dependent on the borrower's future cash-flow expectations (\cite{sufi2009bank, ivashina2010bank, campello2011liquidity, acharya2014credit, brown2021weathering}). In particular, the results highlight how this behavior of credit institutions is conditioned by the credibility of the plan and uncertainty of their forecasts and, therefore, confirm that ``idiosyncratic" volatility of cash-flow is the determining factor in estimating the risk of default, especially for high-indebted firms (\cite{campbell2008search}). Finally, the results depend crucially on $\mu$ and $\sigma^2$, in agreement with the studies that demonstrate the importance of the subjectivity and analytical capacity of the financial operators (\cite{crane2020skilled}) and their influence on the cost of debt (\cite{fracassi2016does}).

\begin{figure}[]
  \centering
\makebox[\textwidth][c]{
\subfloat[] 
	{	\includegraphics[scale=0.5, trim=30mm 80mm 30mm 80mm\textwidth]{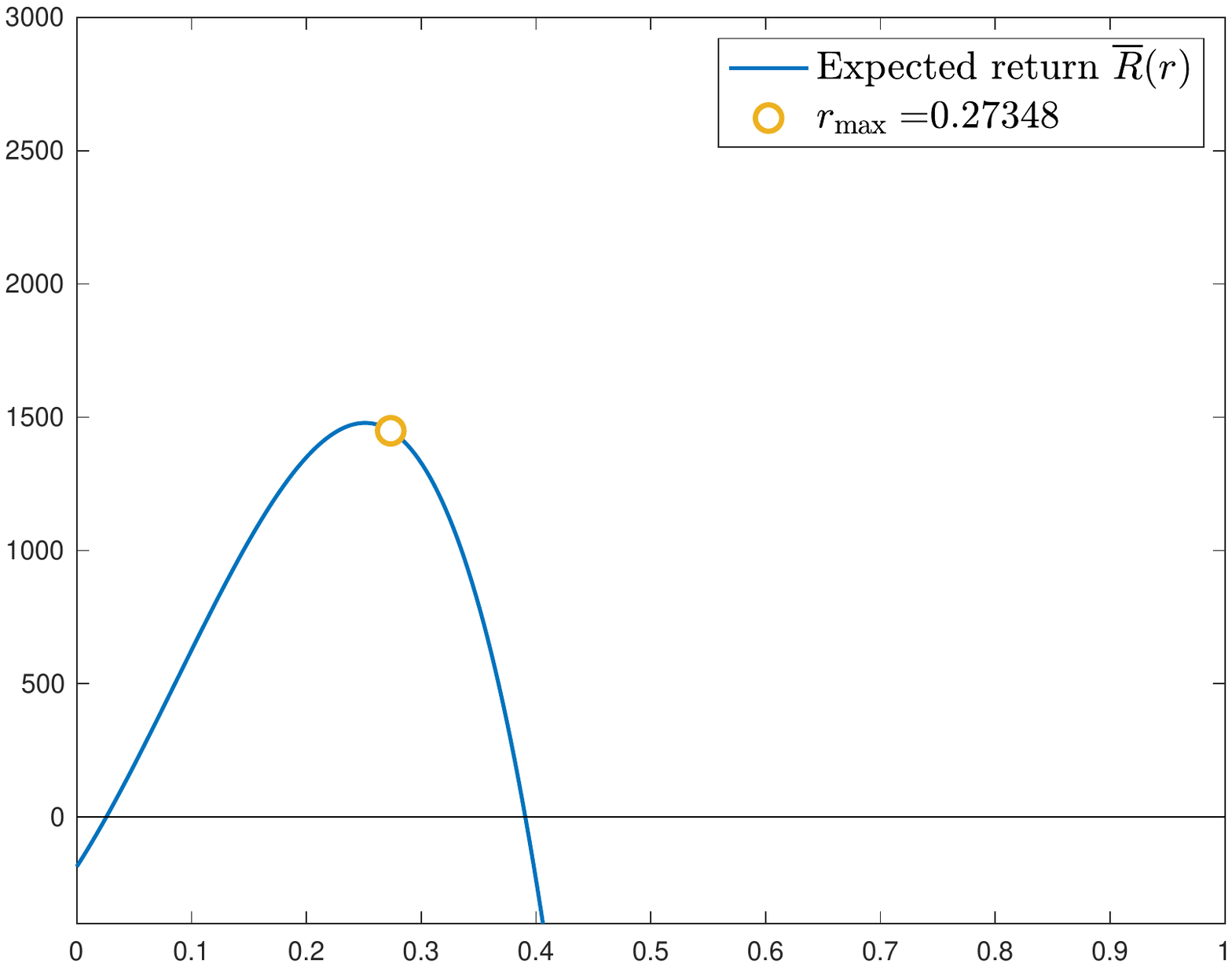} \label{fig7a}} 
	\hspace{1.4cm}
\subfloat[]
	{	\includegraphics[scale=0.5, trim=30mm 80mm 30mm 80mm\textwidth]{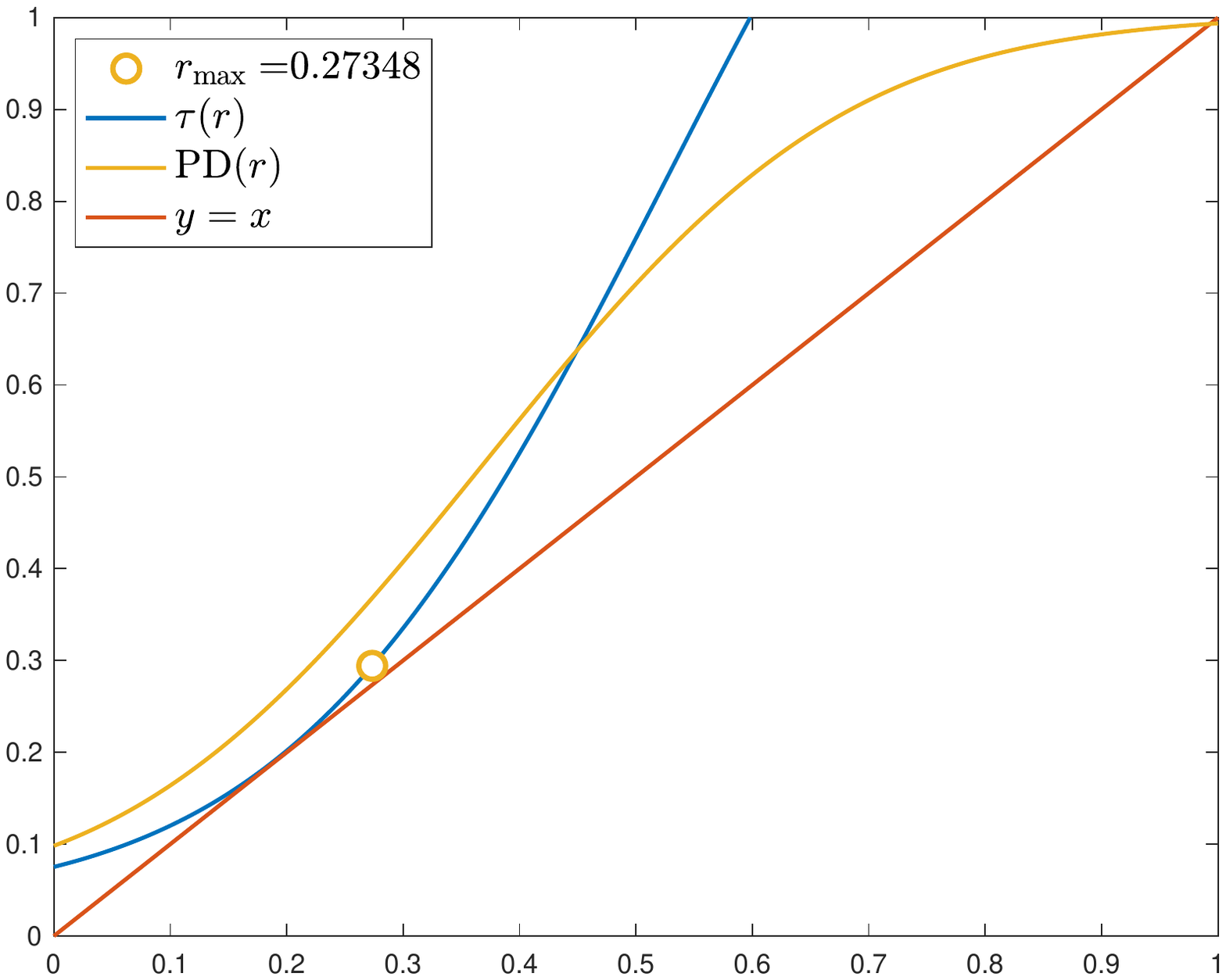} \label{fig7b}}
	} 
\\
\subfloat[]
	{	\includegraphics[scale=0.7, trim=0mm 0mm 0mm 0mm\textwidth]{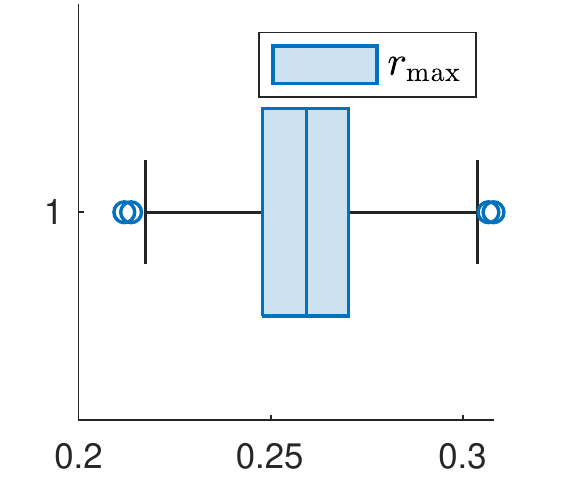} \label{fig7c}}
\caption[]{Simulation results for Case A with the maximum STNFP sustainable.  Figure~\ref{fig7b}: there is only one fixed point of $\tau$, which corresponds to the tangent point of $\tau$ and the $y=x$ curve. Hence, the theory dictates that $r_{\min}$, $r_{\mathrm{fix}}$ and $r_{\max}$ should all coincide. The reason $r_{\max}$ displayed in the plot does not correspond to the tangent point is due to numerical inaccuracies.} 
\label{fig7}
\end{figure}

	To respond to the fourth question posed in the introduction, we determine the highest initial debt sustainable by the company by means of a ``trial and error" method. We perform our analysis repeatedly for increasing choices of the initial debt. As the initial STNFP grows, the equilibrium rate curve rises until it becomes the tangent of the $y=x$ curve. When the two curves are tangent, there is only one equilibrium rate ($r_{\min}$ corresponding to $r_{\mathrm{fix}}$). 
	Figure~\ref{fig7} shows the results for Case A. The maximum initial debt that would be sustainable at a rate of 23.429\% is equal to $d_{S,0}=3,211.11$. Beyond this threshold, equilibrium does not exist.

Finally, to answer the fifth and final question posed in the introduction, we first assumed a reduction to 5 years in the maturity of the term debt compared to the 10-year base scenario. Consistent with the finance theory, the borrower's financial risk should rise, considering the big increase in the principal payments. In fact, the results show that the minimum interest rate increases by 0.014024 as compared to Case A (see Figure~\ref{fig4} vs Table~\ref{table4}).

  \begin{table}[ht]
\centering
\caption{Equilibrium rates in Case A with term debt maturity of 5 years.}
\label{table4}
\begin{tabular}{|l|c|c|}
\hline
 $r_{\min}=0.0786$ & $r_{\mathrm{fix}}=0.8852$ & $r_{\max} = 0.3500$ \\
\hline
\end{tabular}
\end{table}

On the data in question, we carried out a further experiment, again aimed at verifying the effects of debt restructuring maneuvers. In particular, following the contraction in the maturity of the term debt, we hypothesize that the borrower is able to provide collateral in order to reduce the cost of its debt. Assuming that the LGD is halved following this move, the equilibrium rate descends, in line with expectations, and even results lower than the starting rate (see Figure~\ref{fig4} vs Table~\ref{table5}).

  \begin{table}[ht]
\centering
\caption{Equilibrium rates in Case A with term debt maturity of 5 years.}
\label{table5}
\begin{tabular}{|l|c|c|}
\hline
 $r_{\min}=0.0598$ & $r_{\mathrm{fix}}=$n.a. & $r_{\max} = 0.3211$ \\
\hline
\end{tabular}
\end{table}

Also, these results are confirmed in the literature and offer an interpretative line to its results. First of all, they support the recent literature that notes how debt structure and maturity have a significant impact on corporate dynamics and debt overhang (\cite{diamond2014theory, demarzo2021leverage}), contrary to historical theories (\cite{merton1974pricing, leland1998agency}). In this way, our model is aligned with \cite{campbell2021structuring}, showing the harmful impacts of term loans with short maturity period, especially for distressed companies. Finally, the results are aligned with (and explain) empirical research showing the impact of collaterals on the cost of debt (\cite{cerqueiro2016collateralization, benmelech2022secured}).

\section{Conclusion} \label{conclusion}

A company is a forward-looking competitive system and is considered in equilibrium as long as its stakeholders have trust in that future. In moments of foreseeable crisis in that system, credit institutions are the stakeholder who ultimately decides whether, to what extent and until what point to sustain this equilibrium. Institutions will continue to renew and increase a company's lines of credit as long as they consider it profitable. Essentially, this depends on the probability that they attribute to the irreversible growth of the firm's credit lines in a steady state. This probability of default is measured by the credibility and uncertainty of the business plan and updated on the basis of interest rates sustainable by the plan itself. Therefore, equilibrium is a situation of stable and permanent meeting points between the predictable trends of the supply and demand of credit.

On the basis of this theory, in a coherent probabilistic environment we model the borrowing company's equilibrium based on the foreseeable conditions of its credit demand (business plan, LGD, etc.) and on the foreseeable conditions of its credit supply segment (interest rate function, availability of information and analysis tools, etc.). These predictions influence each other. The model quantifies the PD by estimating the intensity of verification of the future default event. This PD is a unique numerical estimate (which is why the interest rate regime between the various forms of debt depends on factors external to the PD). This PD is a function of the rate applicable in the future by credit institutions and vice versa.

In response to the research questions, the model provides important results: i) given an interest rate to be applied on the debt, it provides the probability that the company will remain viable in the future; ii) it verifies the existence of a rate capable of ensuring company's financial health and simultaneous minimum satisfaction of the lenders; iii) it verifies the existence of a rate that maximizes the profitability of lenders, while ensuring company's financial health; iv) it estimates the intensity of the negotiating strength of the borrower and lender; v) it determines the maximum level of sustainable debt at rates deemed satisfactory for the lenders (if it exists); vi) it determines the impact on corporate health of a certain debt structure/restructuring. This default theory and the related model can uniform studies and instruments in different fields (corporate finance, credit risk management, financial intermediation, structured finance, project finance, corporate restructuring, etc.).

An operational outcome of the model is the creation of a ``tailored" scoring system on the debtor's financial market, where competitive forces, management credibility, business development, analytical skills, availability of information, rate curves, and endogenous and exogenous incentives of the supply market interact dynamically. A model emerges that has the following characteristics.

First, it makes extensive use of soft information based on the future of the company, which has been shown to be fundamental in assessing credit risk. There are two advantages: first of all, the model can also be applied to start-ups, companies undergoing restructuring or in radical transformation, and second of all, any foreseeable internal (strategic-operational) and external (competitive and credit supply) changes to the company can be subject to an evaluation update with a real-time response. Because of this capability, generally the model can be used by credit institutions in the granting-renewal-monitoring phases and also in the pricing phase, as well as by market operators to valuate and price bonds.

Second, human assessment skills are integrated into the scoring model, as literature and regulatory bodies have long recommended. The introduction of human subjectivity in the models reproduces the complexity of the market, valorizing operator know-how and freedom\footnote{``Credit risk analysis is an art as well as a science. It is a science because the analysis is based upon established principles emanating from a body of knowledge and sound logic. Individual skill and the way the principles are applied constitute the art element," \cite{joseph2013advanced}. }.  This benefits the entrepreneurial innovative push, as well as a healthy competitiveness among financial operators.

Third, the assessment is centered on the evolution of the borrower's debt structure, taking into proper consideration the debt maturity, rate variability, collateral, etc. This makes it possible to quantify the debt service for each year and to arrive at an accurate estimate of the PD. Consequently, the mechanisms should benefit for selecting firms, structuring the debt based on the characteristics and duration of their financial needs, setting covenants, and fixing the correct price to each type of facility (``the optimal commitment contract is conditioned on borrower-specific variables," \cite{shockley1997bank}). This should improve the functioning of the financial market, mitigating the misallocation of financial liabilities (\cite{campbell2021structuring, whited2021misallocation}).

Fourth, the forecast of default is articulated up to the achievement of a steady state. Consequently, the PD estimate horizon is generally extended compared to lag models, which show myopic predictive capabilities. The advantage of this is the ability to evaluate the real sustainability of the company's business model, especially in terms of socio-environmental durability, an increasingly important condition.

Read jointly, these characteristics of the model allow for the introduction of idiosyncratic risk assessment in scoring models, taking into account interaction mechanisms between incentive systems specific of each operator. This gives rise to a truly forward-looking scoring model, which, as such, should not be affected by the classic stationary limits of today's most widespread scoring systems which have generated serious market inefficiencies.

From an operational point of view, our model appears to be aligned with: a) the definition of default for the programming of credit policies (Unlikeliness to Pay); b) the logic of international regulators who push toward forward-looking models with increasing attention to the Debt Service on the part of operating cash flows (\cite{european2020guidelines}); c) the criteria for restructuring non-performing exposures, helping to fix maturities, rates and covenants in line with the foreseeable evolution of cash flows (\cite{european2018guidelines}); d) the forward-looking credit measurement criteria set by IFRS 9\footnote{The comparison between the rate applied in the credit relationship and the minimum rate determined by the model could help to quantify the fair value of the financial instrument.}  on the basis of which the regulatory capital of the institution is determined and, therefore, also its credit policies; e) the stress testing and scenario analysis criteria, which can be conducted at the level of a single position by acting directly on the revision of the plan's assumptions.

Hence our main conclusion is that the credit risk measurement tools, and the operators who use them, must take a step back in order to move forward, regaining possession of the technicalities of fundamental analysis. It is a question of creating rating systems that ``go back to the future".
Looking at future research prospects, if the risk of default thus determined were injected into the cost of equity, the work could outline interesting perspectives in the field of corporate valuation and in the study of leverage dynamics, a subject far from having reached undisputed results (\cite{demarzo2021leverage}).

Regarding the limitations of the model, we focus on optimal rate strategies with functions that are constant over time (i.e., constant rate\footnote{See note~\ref{note26}.}).  Given that at this time, it is not clear to us how to extend the fixed-point analysis to dynamic rate strategies over time, this problem constitutes an interesting direction of future research.

\appendix
\section{Default}

\begin{prop} \label{appendix}
If the STNFP is increasing (resp. strictly increasing) at some time $\bar t\geq t_{SS}$, that is 
\begin{equation}
D_{S,\bar t}\geq D_{S,\bar t-1}, 
\end{equation}
then the STNFP is increasing (resp. strictly increasing) at each time $t$ after $\bar t$, that is 
\begin{equation}
D_{S,t} \geq D_{S,t-1},\forall t\geq \bar t.
\end{equation}
\begin{proof}
For every $t>0$ it holds $D_{S,t}=D_{S,t-1}-C_t$. Hence,
\begin{equation}
D_{S,t}\geq D_{S,t-1} \Leftrightarrow C_t \leq0.
\end{equation}
Moreover, $C_t$ can be written as $C_t=F_t-K-I_{S,t}$, where $K=c_t+I_{L,t}$ is constant for all $t$. If $t>t_{SS}$, then by assumption $F_t=F$ is constant. Let us consider a fixed time $\bar t \geq t_{SS}$, such that $D_{S,\bar t}\geq D_{S,\bar t-1}$. From our discussion above, it follows that
\begin{equation}
C_{\bar t} = F-K-I_{S,\bar t} \leq 0.
\end{equation}
Since at time $\bar t+1$, $I_{S,\bar t+1}=\mathbf{r} D_{S,\bar t} \geq \mathbf{r} D_{S,\bar t-1}=I_{S,\bar t}$, we have
\begin{equation}
0\geq C_{\bar t}=F-K-I_{S,\bar t}\geq F-K-I_{S,\bar t+1}=C_{\bar t+1}.
\end{equation}
Since $C_{\bar t+1}\leq 0$, it follows that $D_{S,\bar t+1}\geq D_{S,\bar t}$. \\
We showed that, if $D_{S,\bar t}\geq D_{S,\bar t-1}$ for $\bar t\geq t_{SS}$, then $D_{S,\bar t+1}\geq D_{S,\bar t}$. By iterating this computation, the conclusion follows. Note that if $D_{S,\bar t}>D_{S,\bar t-1}$, then the same chain of inequalities allows us to conclude that $D_{S,t}>D_{S,t-1}$, $\forall t\geq \bar t$.
\end{proof}
\end{prop}

Because each implication of the proof is an equivalence, we also have the following corollary:
\begin{cor}
If the STNFP is decreasing at some time $\bar t\geq t_{SS}$, that is $D_{S,\bar t}<D_{S,\bar t-1}$, then the STNFP is decreasing at each time $t$ after $\bar t$, that is $D_{S,t}<D_{S,t-1}$, $\forall t\geq \bar t$.
\end{cor}

Thanks to these propositions, we can claim that the company defaults if and only if right after the steady state the STNFP will not decrease, i.e., $C_{t_{SS}+1}\leq0$.


%
%
%



\bibliographystyle{apalike} 
\bibliography{mybib} 




\end{document}